\documentclass[%
 reprint,
amsmath,amssymb,aps,
pra,
]{revtex4-2}
\usepackage[T1]{fontenc} % add this line to support \DJ in T1 encoding
\usepackage{xcolor}
\usepackage{graphicx}% Include figure files
\usepackage{dcolumn}% Align table columns on decimal point
\usepackage{bm}% bold math
\usepackage{hyperref}% add hypertext capabilities
\usepackage[mathlines]{lineno}% Enable numbering of text and display math
\usepackage{booktabs}

\begin{document}

\preprint{}

\title{Overcoming the Thermal-Noise Limit of Microwave Measurements\\
by Pre-cooling with an Active Cold Load}% Force line breaks with 

\author{Kuan-Cheng Chen}
 %\altaffiliation[Also at ]{Department of Materials, Imperial College London, South Kensington SW7 2AZ, London,
%United Kingdom}%Lines break automatically or can be forced with \\
\author{Mark Oxborrow}%
 \email{m.oxborrow@imperial.ac.uk}
\affiliation{%
 Department of Materials, Imperial College London, South Kensington SW7 2AZ, London,
United Kingdom 
}%

\date{\today}

\begin{abstract}
We introduce a method, which we here name ``active pre-cooling'' (APC), for removing, just prior to performing a measurement, a large fraction of the deleterious thermal photons that would otherwise occupy the electromagnetic modes of a microwave cavity or some alternative form of radio-frequency resonator.
The removal is achieved by temporarily over-coupling the cavity's modes to an active cold load (ACL). We report a room-temperature bench-top demonstration of the method, where this load takes the form of the input of a commercial low-noise amplifier (LNA). No isolator is inserted between the LNA's input and the cavity's coupling port. 
The noise temperature of a monitored microwave mode drops to $\sim$123~K. Upon incorporating our pre-coolable cavity into a time-resolved (tr-)~EPR spectrometer, a commensurate improvement in the signal-to-noise ratio is observed, corresponding to a factor-of-5 speed up over a conventional tr-EPR measurement at room temperature for the same precision and/or sensitivity.
Modeling indicates the feasibility, for realistic mode quality factors and couplings, of cooling the room-temperature cavity's modes down to a few tens of K, and for this coldness to last several tens of \(\mu s\), whilst the cavity and its contents are optimally interrogated by a microwave tone or pulse sequence.
The method thus provides a simple and generally applicable approach to improving the sensitivity and/or read-out speed in pulsed and time-resolved EPR spectroscopy, quantum detection and other radiometric measurements. 
It provides a first-stage cold reservoir (of microwave photons) for other, deeper cooling methods to work from and, through the use of cryogenic ACLs (realized, in the first instance,
as the inputs of existing cryogenic low-noise amplifiers),
the method could itself be directly extended to lower temperature regimes.

%\begin{description}
%\end{description}

\end{abstract}

\maketitle

%\tableofcontents

%\section{\label{sec:level1}First-level heading:\protect\\ The line
%break was forced \lowercase{via} \textbackslash\textbackslash}

\section{Introduction}

In the face of thermal noise, quantum measurements very often require cryogenic environments to achieve sufficient sensitivity and/or fidelity. In the microwave regime of frequencies, this involves cooling the hardware's cavities and connecting waveguides down to a temperature of a few tens of millikelvin (mK) so as to reliably eliminate the incoherent thermal photons that would otherwise reside within them and cause havoc. Obeying Bose-Einstein statistics, the expected number of such photons occupying any one of the cavity's electromagnetic (EM) modes is given by $\bar{n} = [e^{(hf_{\text{mode}}/k_{\rm B} T})-1]^{-1}$, where $f_{\text{mode}}$ is the mode's frequency. This level of cooling can be achieved, albeit onerously, by installing the hardware within a deeply cryogenic refrigerator, most often a dilution or adiabatic-demagnetization fridge.

Though currently still a few factors of ten greater in temperature/photon number than what can be achieved through standard mK refrigeration (as involves the removal of thermal phonons from the materials that surround or fill an electromagnetic cavity),
recent studies\cite{wu2021bench,ng2021quasi,fahey23,wang-spin-refridge24,day24} have demonstrated the direct removal of thermal photons from particular microwave modes inside bench-top cavities at room temperature through the stimulated absorption of these photons by an optically spin-polarised medium placed inside the cavity.
While \mbox{obviating} the need for physical refrigeration, this hybrid (optical-microwave) approach still requires a sufficiently powerful  optical pump source for spin-polarizing the absorbing medium. This results in still quite a bulky and power-hungry system, even when a miniaturized microwave cavity of high magnetic Purcell factor\cite{breeze15,fleury23} is used to reduce the optical pump power/energy needed.

Within microwave instrumentation, electrically powered low-noise amplifiers (LNAs) are ubiquitously
used to increase the amplitude (and thus power) of weak signals. Guided quantitatively by Friis’ formula\cite{friis44}, the deployment of the quietest LNA available at the front end of an instrument's amplification chain will minimize the amount of deleterious noise added to the measured signal, thereby maximize the overall signal-to-noise ratio (SNR) and measurement sensitivity\cite{vsimenas2021sensitivity}. 
Compared to their manifest utility in radio-astronomy\cite{giordmaine1959maser} and deep-space communication\cite{clauss2008}, the gains in sensitivity from installing a lower-noise (cryogenic) preamplifier into an electron paramagnetic resonance (EPR) or nuclear magnetic resonance (NMR) spectrometer are found, empirically, to be both (i)~modest\cite{collier68,narkowicz13,kalendra22}, especially when the spectrometer's microwave or radio-frequency (RF) cavity is not itself refrigerated, and (ii)~dependent on the degree of cavity-amplifier coupling\cite{pfenninger95}.

Various schemes such as Townes' ``\textit{Q}-multiplier''\cite{townes1960sensitivity}, involving the incorporation of a maser gain medium inside the cavity itself, have been proposed over the years towards improving measurement sensitivity. It was \mbox{Mollier}
\textit{et al.}\cite{mollier1973theoretical} who cut through the confusion to affirm that, no matter how low-noise the receiving LNA gain medium, the sensitivity of an EPR (or NMR) spectrometer is ultimately limited by thermal noise (\textit{i.e.}~incoherent microwave photons) generated from within the EPR cavity itself. An ultra-low-noise receiving amplifier attached directly to the cavity's coupling port, or even an ultra-low-noise gain medium resident inside the cavity, will amplify both the signal and this noise equally; thus, the SNR won't improve. As addressed quantitatively by Poole\cite{poole2019electron} and
others\cite{hill1968limits,rinard1999frequency} since, the only ways to improve sensitivity (for a given sample, completely filling the cavity, at a given temperature) are to either increase the cavity's quality factor $Q$ (albeit at the expense of measurement bandwidth), so increasing the photon cavity lifetime, or to increase the EPR transition frequency (by applying a strong d.c. magnetic field to the sample), so reducing the number of thermal photons in the cavity's operational mode. Or else (back to where this argument began:) to physically cool the cavity or otherwise ``state-prepare'' its operational microwave mode(s).

At least for frequencies in the low-GHz region, today's solid-state LNAs,
in the form of monolithic microwave integrated circuits (MMICs) or else as hybrids composed of interconnected discrete components, containing either InP- or GaAs-based
high-electron-mobility transistors (HEMTs) or else SiGe-based heterojunction bipolar transistors (HBTs), offer impressively low noise temperatures (referred to input) of 50~K or less --even when operating at room temperature. As mature commercial-off-the-shelf (COTS) components, they are readily purchasable. If suitably designed/chosen, the noise temperature of such an LNA will decrease by a further order of magnitude or so upon cooling to liquid-helium temperature\cite{weinreb1982ultra,pospieszalski05}. With such low amplifier noise temperatures available, the sensitivity with which the field-amplitude/energy ($=$ number of photons) inside a room-temperature microwave cavity can be measured is determined almost entirely by the cavity's own thermal noise (as is associated with the rate at which electromagnetic energy is dissipated inside the cavity) --not the small amount of extra noise imparted by the receiving amplifier chain used to monitor the cavity's field amplitude through a port attached to the cavity.

As already mentioned above, Wu \textit{et al.}\cite{wu2021bench} demonstrated the cooling of an electromagnetic (EM) mode by coupling it to a ``spin-cold'' reservoir in the form of an optically excited crystal of pentacene-doped \textit{para}-terphenyl (Pc:PTP).  
Ng~\textit{et al.}\cite{ng2021quasi}, Blank~\textit{et al.}\cite{blank23}, and Day~\textit{et al.}\cite{day24}, have since demonstrated quasi-continuous mode cooling using photoexcited negatively charged nitrogen vacancies (NV$^-$) in diamond as a spin-polarised absorber. 
In a similar vein, Englund, Trusheim and co-workers\cite{fahey23,wang-spin-refridge24} have exploited
mode cooling with NV$^-$ diamond to enhance the sensitivity of a magnetometer.
All of these works use a cm-sized ``3-D'' dielectrically-loaded cavity to support a TE\(_{01\delta}\) mode affording a high magnetic Purcell factor\cite{breeze15}. The frequency \(f_0\) of this mode is adjusted, or a d.c.~magnetic field is applied, to align it with the absorptive paramagnetic transition being targeted. Since the linewidth of this transition is narrow (a few MHz for pentacene, a few hundreds of kHz for NV$^-$ diamond),  these approaches to mode cooling are intrinsically narrow-band.

\section{Active Cold Loads}
In microwave radiometry, cooled coaxial or waveguide terminations, also known as ``cold loads'', are used both as sources of noise
and as absorbers\cite{gervasi1995absolute} of unwanted signals. Such a cryogenic load is a passive component requiring no source of power.
To be useful as an accurately quantified noise source or ``reference'',
it must be anchored thermally to a heat bath (such as a ``cold plate'' within a cryostat) whose temperature is accurately known.
A cryogenic load can be connected to one of the ports of a (cryogenic) circulator to form an isolator. At least within the circulator's operational bandwidth, this isolator can be used to protect a sensitive quantum circuit from backwardly propagating thermal noise and/or coherent signals unintentionally generated by reflections further up/along the signal path.
In this context, we point out that the pre-cooling method introduced here draws parallels with the radiative cooling of a spin ensemble or resonator, held physically at liquid-helium temperature, by coupling it electromagnetically to an absorber (in the form of either a passive load or active device) held at mK temperature\cite{albanese2020radiative,wang21}.

\begin{figure}[htpb]
\includegraphics[width=0.9\columnwidth]{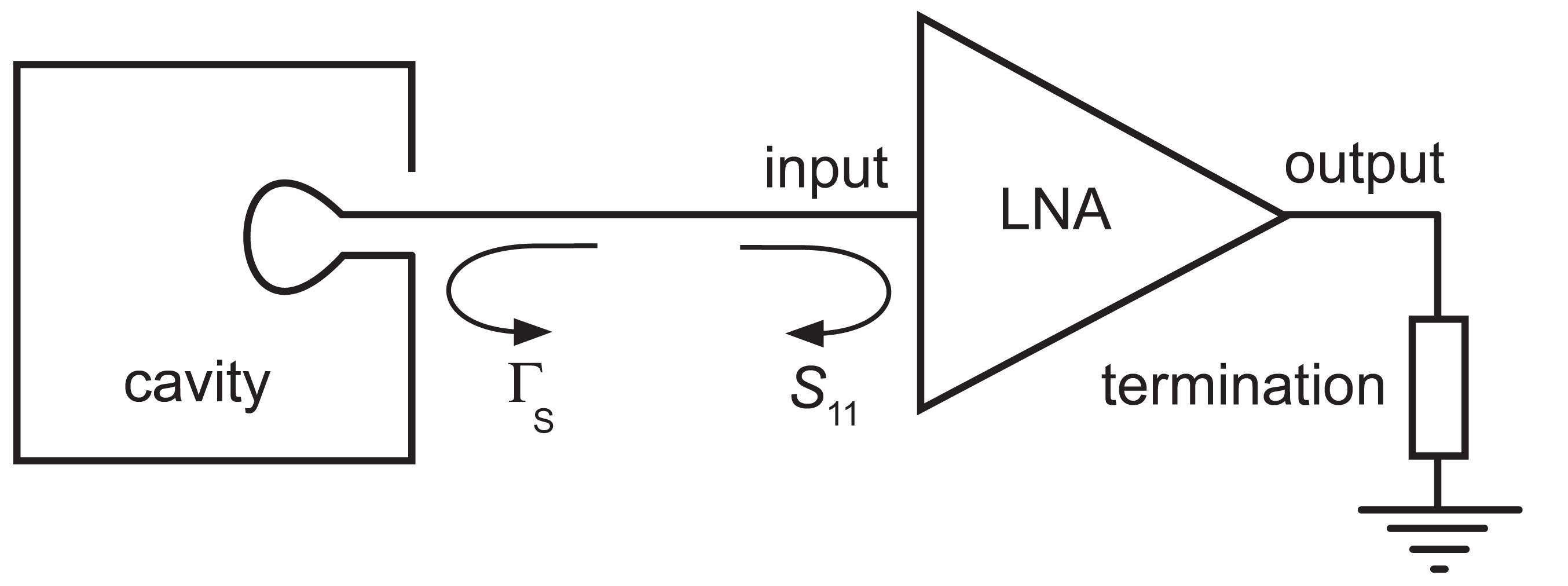}
\caption{\label{fig:acl_connected}
Paradigm of a cavity that is externally coupled to an active cold load (ACL) in the form of the input of a low-noise amplifier (LNA) whose output in terminated.}
\end{figure}

A convenient alternative to a physically cryogenic termination is an \textit{active} cryogenic load (ACL)\cite{frater1981active,forward1983active}, which, though a powered device, mimics the former's behavior. ACLs can be made from the inputs (gates/bases) of suitably biased low-noise transistors
configured as single-port terminated amplifiers\cite{dunleavy1997,jarrige2010active,weissbrodt2017}. 
Though not always cold enough (nor broad-band or stable enough) to replace truly cryogenic terminations in all applications,
they can be used advantageously in airborne and space applications to calibrate (or at least drift-correct) earth-monitoring radiometers\cite{shou2015,weissbrodt2017}, dispensing with the encumbrance of mobile refrigeration and/or the operational need to “roll” the instrument's platform to point the radiometer at dark sky (\textit{i.e.}, at 2.7~K cosmic background radiation --serving as an absolute temperature reference).

Acknowledging that other approaches exist\cite{clerk2010introduction}, we here apply the so-called
``wave approach'' to analyse the generation and flow of noise in realistic microwave circuits (see chapter 8 of Ref.~\cite{siegman1964microwave}, or Refs.~\cite{meys1978,djordjevic2017wave}).
Through the construction of flow graphs combined with the application
of Mason's ``non-touching loop'' rule\cite{hpAN154,hecken1981}, 
this diagrammatic approach readily encompasses the frequently encountered
complexity of what a laser physicist would describe as ``etalon effects'', 
where simultaneously forward- and back-propagating noise waves form (partial) standing waves
along sections of transmission line between two or more impedance mismatches (the equivalent of partially
reflecting mirrors functioning as beamsplitters/combiners).  
The \textit{available} noise temperature from the input of a low-noise amplifier (of sufficiently high gain) that is terminated by
a matched and sufficiently cold load can thereby be quantified as\cite{meys1978,escotte2012}:
 \begin{equation}
    \label{lnaIPnoise}
   T_{\text{cold}} = -T_\text{min} +
   4 T_0 \frac{R_\text{n}}{Z_0} \frac{|1- S_{11} \Gamma_\text{opt}|^2} {(1-|S_{11}|^2) |1 + \Gamma_\text{opt}|^2},
\end{equation}
where \(S_{11}\) is the (in general complex) scattering parameter for reflection off the LNA's input and where 
\( T_\text{min}\)~\([\text{K}] \), \( R_\text{n}\)~\( [\Omega ] \), and \( \Gamma_\text{opt}\)~\( [\text{dimensionless, complex}] \) compose the LNA's set of industry-standard noise parameters\cite{dunleavy2009,modelithicsAN-60-040} (the associated unit of each is given in square brackets),  and where \(T_0\) and \(Z_0\) are the reference temperature and line impedance associated with the definitions of same.
For an LNA intended to operate at room temperature, that is 
constructed and noise-calibrated using 
the most common types of waveguide/cable and connector at low microwave frequencies,
its is likely that \(Z_0 =\)~50~\( \Omega\), and  \(T_0 \)~=~290~K for the noise parameters provided
(if one is lucky) by the LNA's manufacturer. 
At the cost of more complicated expressions (incorporating contributions from additional paths/loops according to Mason's rule,
thereby involving  more of the LNA's scattering parameters), the above formula can be extended to include the effects of a both
warm and non-matched (thus partially reflecting) termination at the LNA's output\cite{wait1991,weatherspoon2006, escotte2012}.

Note that the minus sign in front of \( T_\text{min} \) in Eq.~\ref{lnaIPnoise}
can lead to (partial) cancellation between the two terms on the r.h.s.~of the same, 
all depending on the exact magnitude of \( R_\text{n} \) and the locations of   
\( S_{11} \)  and \( \Gamma_\text{opt} \)  in the complex plane. 
For the particular LNA used in our room-temperature experiments reported here, namely a Qorvo QPL9547, we calculate, from its specification sheet\cite{QPL9547}, that \(T_{\text{cold}}~=~29.1\)~K at 1.45 GHz --around 10~times lower than the LNA's operating temperature. 

Alas, complete sets of scattering and noise parameters are rarely published for cryogenic amplifiers, preventing immediate off-paper predictions of their performance when repurposed as cryo-ACLs. One informative exception is the noise calibration, exploiting a cryogenic impedance tuner, of the type of LNA used for the ``ALPACA'' phased-array radio telescope, as reported in ref.~\cite{sheldon2021}. Though only the magnitude of \( S_{11} \) (not its angle) is provided, upon reading off from Figs.~12 and 13, \textit{ibid.}, one can frame this LNA's available input noise temperature to within: 6.2~K~\( < T_{\text{rev}} < \)~7.3~K at 1.5 GHz, when operating at 20~K. 
Equivalent characterizations (including complex scattering parameters) 
of transistor-based, maser, and parametric amplifiers operating at even lower temperatures would
enable the rational design of improved cryo-ACLs.   

\begin{figure}[b]
\includegraphics[width=\columnwidth]{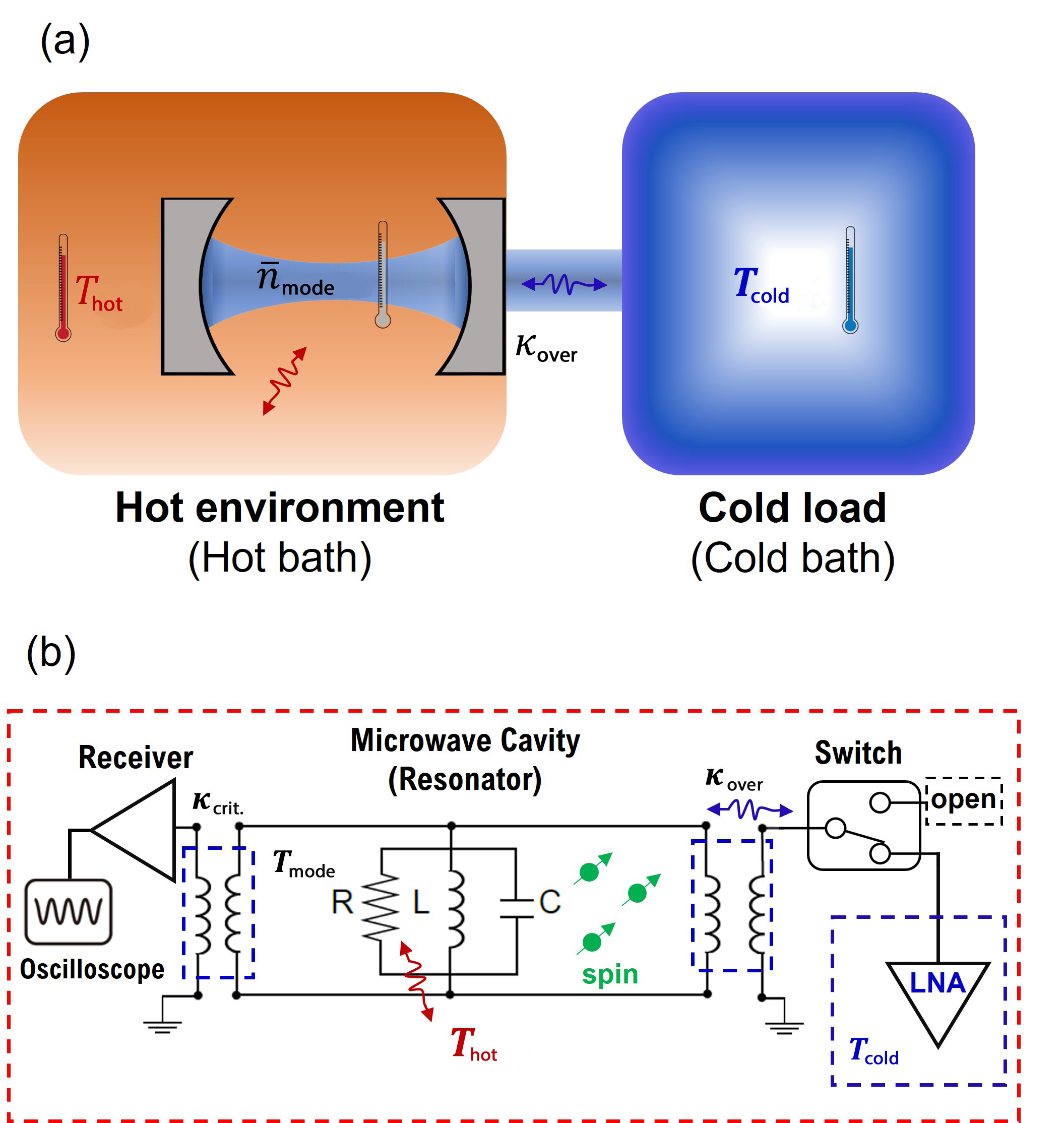}
\caption{\label{fig:fig1}(a) Conceptual mechanism of active pre-cooling (APC):
A microwave resonator is physically maintained at a temperature \(T_{\text{hot}}\)
(in this article, \(T_{\text{hot}}\) equals room temperature, denoted as \(T_\text{0}\)) and supports an 
electromagnetic mode whose photon occupation number is \( \bar{n}_{\text{mode}}\). This mode is coupled to a cold load maintained at a lower temperature \(T_{\text{cold}}\)
with a coupling factor of \( \kappa_{\text{over}} \) between the load and the mode. 
(b) Simplified equivalent electrical circuit: The resonator, here drawn as an LCR tank circuit, is over-coupled (right) through a (SPDT) switch to either a cold load (in the form of the input of an LNA) or an open circuit --depending on the position of the switch.
The mode's field amplitude (as generates a voltage across the tank circuit) can be measured by a second LNA through its own coupling port (left --this port's coupling is usually adjusted to be at critical coupling when the cooling LNA is switched out) whose output is recorded on a digital oscilloscope.}
\end{figure}

Finally, it is instructive to compare Eq.~\ref{lnaIPnoise} against the standard expression for the effective noise
temperature of an LNA\cite{wait1991,dunleavy2009,modelithicsAN-60-040} when used, as intended, as an amplifier:
 \begin{equation}
    \label{lnaOPeffnoisetemp}
   T_{\text{eff.}} = T_\text{min} + 4 T_0 \frac{R_\text{n}}{Z_0}
   \frac{(|\Gamma_\text{s} -|\Gamma_\text{opt}|^2)}
   {|1+ \Gamma_\text{opt}|^2
   (1-|\Gamma_\text{s}|^2)},
\end{equation}
where \( \Gamma_\text{s} \) is the coefficient of reflection off the source connected to the input
of the LNA --in our case the port of a microwave cavity (see Fig.~\ref{fig:acl_connected}), possibly with a matching circuit/stub interposed. This expression will be needed later on for determining the temperature of a microwave mode --see Supplemental Material\cite{supplemental} (which includes references \cite{escotte1993evaluation,sung2003transient,hassan2010reducing}). \(T_{\text{eff.}} \) can be  minimized by ``noise matching'', \textit{i.e.}~adjusting  \( \Gamma_\text{s} \) to equal \( \Gamma_\text{opt} \) in the complex plane (\textit{viz.}~on the Smith chart). Note that minimizing \(T_{\text{cold}} \) for an active cold load constructed from an LNA is a different task (their respective formulae \ref{lnaIPnoise} and \ref{lnaOPeffnoisetemp} above are different).  
 
\section{Active Pre-Cooling}
The method's essential idea is to first (i) cool the electromagnetic modes of a microwave cavity containing a sample by over-coupling them to an external active cold load (ACL) and then (ii) swiftly reduce this coupling such that the sample can be interrogated in an electromagnetically cold
(and thus quiet) environment before the modes warm back up to ambient temperature. Doing so may enable a more sensitive measurement than otherwise possible. 

The first, cooling step is depicted in Figs.~\ref{fig:acl_connected} and \ref{fig:fig1}(a).
In the bench-top experiments we report in this article, the physical walls and contents of the cavity, and the ACL attached to it, all  lie at room temperature (nominally, 290~K).
Note that the ACL is an electrically powered, single-port device.
The \textit{output} from the LNA that realizes the ACL does not (need to) go anywhere; it can be sunk into a matched passive load at ambient temperature or, if the attenuation backwards through the cooling LNA (essentially \( |S12| \)) is insufficient, into a second ACL --provided by the input of an additional LNA whose output is terminated by an ambient matched load. [We point out parenthetically that the idea of advantageously terminating the output of the ``primary'' cooling LNA with a ``secondary'' cooling LNA (and so on in cascade, as far as needs be)
is not new\cite{lardizabal1997}.]
In our experiments, the input of the cooling LNA is connected via a low-insertion-loss switch to a large loop of wire located inside the cavity, which serves as an over-coupled port.

\begin{figure}[b]
\includegraphics[width=0.75\columnwidth]{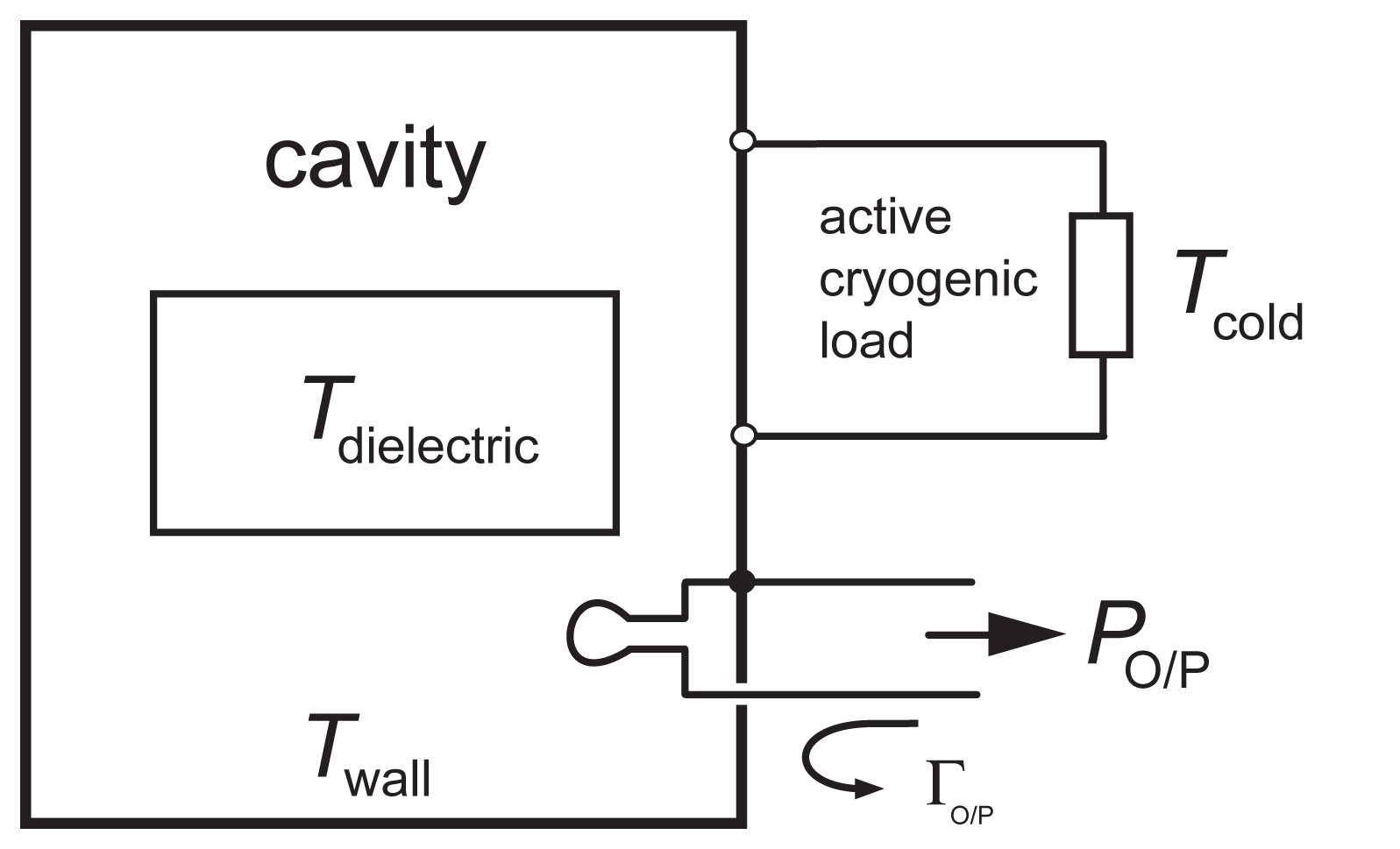}
\caption{\label{fig:Siegman_useful}
The cavity's dissipators, their temperatures, and the resultant noise power outputted.}
\end{figure}

For understanding and quantifying the effects of coupling an external ACL to a microwave cavity, we now appeal to what Siegman
describes simply as a ``useful noise theorem'' in section 8.3 of his book\cite{siegman1964microwave}. 
Consider a cavity whose internal electromagnetic field is monitored by a small (weakly coupled) probe attached to an \mbox{external} port, so providing an output (``O/P''; see Fig.~\ref{fig:Siegman_useful} --the equivalent of Fig.~8-10, \textit{ibid.}). Here, one should regard the cavity's field as the common ``bus'' (or ``loss-less network'' --in Siegman's more circuit-oriented parlance) through which  all of it components, including external loads, interact. 
Applying the theorem: The available noise power per unit bandwidth \( B \) of an electromagnetic mode inside the cavity expressed in units of temperature, namely \(T_{\text{mode}} =  P_{\text{O/P}}  / [k_\text{B} B (1 - | \Gamma_\text{O/P}|^2) ] \),  will equal the weighted sum of the temperatures \( T_i\) of the \( N \) bodies that dissipate the mode's energy, where the weighting  \( p_i\) given to each body equals the fraction of the mode's energy that it dissipates, and where the sum of these fractions adds up to unity: \( \sum_i^N p_i = 1\). Mathematically (\textit{cf.}~equation 8-3-3 in Ref.~\cite{siegman1964microwave}): \(T_{\text{mode}} = \sum_i^N p_i T_i \). We stress that \(T_{\text{mode}}\) quantifies the \textit{available} noise power (per unit bandwidth), taking into account, through the above factor of \( (1 - | \Gamma_\text{O/P}|^2) ] \), the rate at which noise energy is extracted by the probe, where
\( \Gamma_\text{O/P} \) is the reflection coefficient off the probe's port (again, see Fig.~\ref{fig:Siegman_useful}). For the sake of the argument here, we assume that the probe's coupling and thus the noise power the probe itself extracts, as will get sunk into a matched load (at the input of a monitoring instrument), is infinitessimally small, such that this 
load's dissipation does not contribute to the \( \sum_i^N \)~sum.
One must regard the cavity's enclosing metal walls (of finite conductivity), any dielectric or magnetic material (of finite loss tangent) inside the cavity, plus any other components that contribute to the mode's intrinsic loss, as such dissipating bodies. External loads coupled to the cavity, such as an active cryogenic load, also count. The thermodynamic temperature that the electromagnetic mode itself adopts is a dissipation-weighted balance between them all.

Over-coupling to an external cold load  (passive or active) guaranties that the temperature of a microwave mode is dominated by the load's available input noise temperature as opposed to the temperature(s) of the cavity's own intrinsic dissipators. Note that the method affords wide-band cooling: it will cool all of the modes lying within the cold load's bandwidth, provided they are sufficiently  coupled to the load.
Quantitatively, what depth of cooling can one expect? Referring back to Figs.~\ref{fig:fig1}\~(a) and ~\ref{fig:Siegman_useful}: we suppose for simplicity that all dissipating components of the microwave cavity (grey) are kept at the same common physical temperature of \(T_\text{hot} = T_\text{0}\) (\mbox{orangey} brown). Consider a microwave mode within the cavity that is coupled, with a coupling factor of \( \kappa \), to a cold load (blue), whose available temperature is \(T_\text{cold}\).  The dissipation generated in the external load will be \( \kappa \) times that of the mode's own intrinsic (\textit{i.e.}~internal)  dissipation. Applying our implementation of Siegman's useful noise theorem: 
on average, \(\bar{n}_\text{mode}\) thermal photons will occupy
the mode, corresponding to a mode temperature of 
\(T_\text{mode} = h f_\text{mode}/[k_{\rm B} \textrm{ln}(1 + 1 / \bar{n}_\text{mode})]\), where
 \begin{equation}
    \label{eq:1}
   T_{\text{mode}} = \frac{T_\text{0}  + \kappa  T_\text{cold}}{1+\kappa}
\end{equation}
One immediately sees from this equation that, for a high coupling factor \( \kappa = \kappa_\text{over} \gg 1\),  corresponding to a substantially over-coupled port, the microwave cavity's noise temperature will approach that of the cold load, $T_\text{cold}$.
The dynamics of the cooling process is treated around Eqs.~S1 and S2 in the Supplemental Material\cite{supplemental}, whose steady-state solution (S3) agrees with the thermodynamic treatment above. The time scale on which the mode is cooled is set by the mode's as-loaded photon lifetime, namely \( Q_0/[2 \pi f_\text{mode} (1 + \kappa_\text{over})]\).

In the case of an active cold load formed by the (gated) channel of the LNA's input HEMT, the height of this channel is just a few nm (defining a layer of two-dimensional electron gas: a ``2-DEG'') and its lateral dimensions (length and width) are sub-microns\cite{bautista2008}. The cooling capacity of this atto-liter volume of semiconductor is limited. Operationally, one requires that the \mbox{density}, in frequency space, of the electromagnetic modes to be cooled is  sufficiently low such that the noise power provided by the total number of modes within the input HEMT's operating bandwidth is insufficient to saturate/overload its channel. This means that the cavity enclosing and defining the modes must not be too large (to prevent the spectral density of modes becoming too high) nor too hot, and the coupling factors of its modes to the ACL must not be too high. 
As a last resort, saturation could be thwarted by inserting a low-loss band-pass filter between the cooling LNA and the cavity to act as a pre-selector, so coupling the ACL to fewer modes.  
A sense of scale is useful with regards to these concerns, however: The noise power impinging on the 
LNA's input, at a coupling factor of \( \kappa \), from a cavity mode at absolute temperature $T$, equals
\( 
 2 \pi f_{\text{mode}} \kappa k_\text{B} T / Q_0,
\)
where \( k_\text{B} \) is Boltzmann's constant.
For the experimental parameters of the demonstrations reported in this article,
this amounts to -121~dBm per mode. The combined power from the several dozen modes lying within our LNA's frequency range (for the dimensions and dielectric loading of the cavity that we used) is thus still around ten \mbox{orders} of magnitude weaker than our LNA's specified saturation input power (at -1 dB compression). How the LNA's noise parameters themselves depend on the input (noise) power at elevated power levels is seldom reported, though this huge ratio suggests that the  cooling capacity of an ACL constructed from a solid-state LNA enjoys substantial headroom. 

\begin{figure*}[htbp]
\includegraphics[width=2\columnwidth]{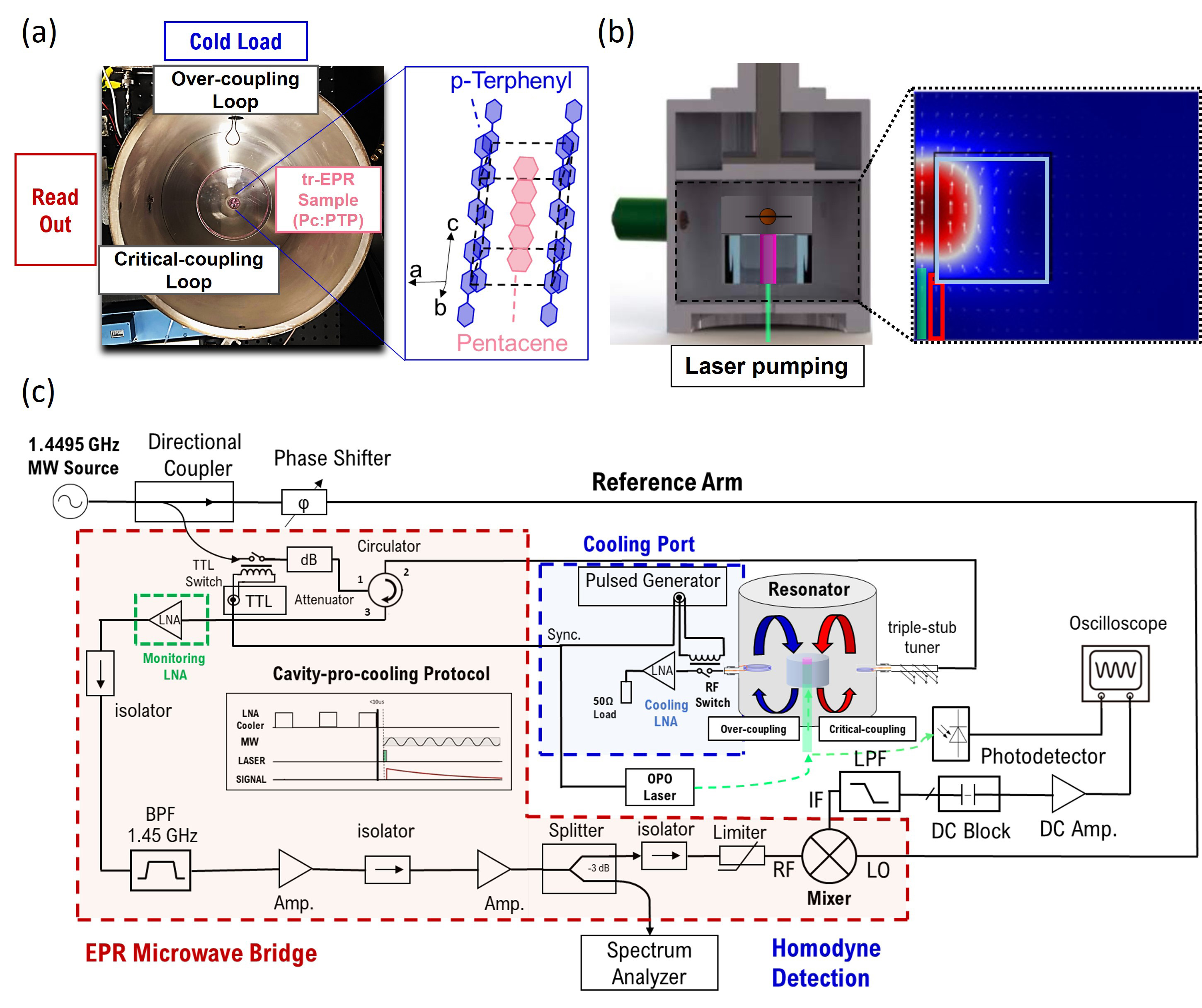}
\caption{\label{fig:fig2} (a) Photograph of the experimental resonator's interior: a sapphire ring holding a 0.1\% Pc:PTP crystal, surrounded by the silver-plated walls of a cylindrical brass enclosure. (b) CAD diagram of the same resonator with a two-dimensional axisymmetric simulation of the magnetic flux (white arrows) and magnetic energy density of the TE\(_{01\delta}\) microwave mode\cite{wu2020invasive} that this resonator supports. (c) Signal-path anatomy of the experimental time-resolved EPR spectrometer used to demonstrate active pre-cooling.
This circuit incorporates homodyne detection of the transient EPR signal. The firing of the optical-parametric oscillator (OPO) laser system and the closing/opening of the two microwave switches are scheduled via transistor–transistor logic (TTL) control pulses (generated by two cascaded pulse generators --both a Thandar TG105) as explained in Fig.~\ref{fig:figStep123}.}
\end{figure*}

\textit{Practicalities (that complicate matters somewhat)}:

[1]~The value of the external coupling, \( \kappa \), associated with a port of the cavity is usually determined by interrogating the port
with a vector network \mbox{analyzer} (VNA) whose connecting leads have a known line impedance --usually 50~\( \Omega \). The input of an ACL that subsequently gets attached to the port will never perfectly match this impedance and so the cavity's coupling to the ACL will be different. One can pragmatically take this into account by defining a  modified coupling: \( \kappa' = \frac{ \text{Re}[Z]}{Z_0} \kappa \), where \(\text{Re}[Z] \) is the real part of the ACL's input impedance \( Z \). Or else one can add a low-loss matching network (equivalent to a series reactance equal to \(-\text{Im}[Z] \)) to the ACL's input. 
The input impedance of the Qorvo QPL9547 LNA (mounted on an evaluation board) at 1.45~GHz that we here repurpose to function
as an ACL is \( Z = (36.5 - \text{i } 18.3)\)~\(\ \Omega \) (from its specification sheet). Its departure from  \( Z = 50\)~\( \Omega \) 
will cause a 27\% drop in the coupling without a compensating network. Also, the partial standing wave (VSWR = 1.35) along the connection between the ACL and the cavity's over coupled port will cause the coupling to vary with the connection's phase length. This can be straightforwardly modeled though,
in our experiments, this length was kept as short as possible (much less than a wavelength at 1.45 GHz) to avoid the complexity.   

[2]~The noise temperature on the same port as seen/felt by the cavity will also be slightly higher than the cooling LNA's (available) input noise
temperature due to insertion losses (I.L.) along this same connection. 
These losses comprise [see Fig.~\ref{fig:fig2}(c)] the I.L.~of the single-pole double-throw (SPDT) switch itself plus those of the two short connecting coaxial cables on either side of it. The increased noise temperature generated by them equals \( \lambda T_\text{0} \), with \( \lambda \equiv L_\text{dB}/4.34 \), \( L_\text{dB} \) being the overall I.L.~in units of dB\cite{siegman1964microwave}; we measured this loss to be 0.19~dB.

[3]~In the experiments reported here, the temperature of a particular (``target'') mode at 1.45~GHz is monitored through an additional second port into the cavity, to which the input of a second LNA is critically coupled (when the cooling LNA on the overcoupled port has been disconnected) --see Step~3 of Fig.~\ref{fig:figStep123}. The output of this monitoring LNA feeds a homodyne receiver tuned to the target mode's center frequency. When modeling the cavity's noise,   
one needs to include the (albeit relatively modest) cooling effect of this critically-coupled monitoring LNA (\textit{viz.}~a second, nominally identical, Qorvo QPL9547EVB-01) that is permanently coupled to the mode.

Including all of the above additional contributions (as per Eq.~S13 of the Supplemental Material\cite{supplemental}), the lowest obtainable mode temperature, \(T_{\text{mode}}^{\text{cooled}}\), is predicted to be
\begin{equation}
     \label{eq:T_EPR}
    T_{\text{mode}}^{\text{cooled}}= \frac{(1 + \kappa'_{\text{over}}\lambda) T_{\text{0}} + \kappa'_{\text{over}} (1 - \lambda) T_{\text{cold}} + T_{\text{mon.}}}{2 + \kappa'_{\text{over}}},
\end{equation}
where $T_{\text{mon.}}$ is the effective cooling temperature of the monitoring LNA taking into account
the unavoidable I.L. of the triple-stub tuner connecting it to the cavity's monitoring port;
this was measured to be 6.05~dB and takes  $T_{\text{mon.}}$ up to 225.2~K. 
The expected noise temperature inside the cavity after pre-cooling,
as calculated from Eq.~\ref{eq:T_EPR} above (see Eq.~S16 in the Supplemental Material\cite{supplemental}),
is $T_{\text{mode}}^{\text{cooled}}$ = 131.5 K.

\section{Switchover followed by Interrogation}

The second essential step in our pre-cooling method is to rapidly disconnect the over-coupled ACL from the cavity without excessively disturbing the pre-cooled mode(s). For the method to work and be useful: (i)~the over-coupling factor, \( \kappa_\text{over} \)  must be sufficiently large to provide a worthwhile depth of cooling, but also (ii)~the switchover between the cavity’s over- and critically-coupled configurations must be sufficiently adiabatic, whilst (iii) the time required to perform the switchover must be a sufficiently small fraction of the cavity’s photon lifetime to leave enough time to exploit the pre-cooling before the cavity re-thermalizes. 
Obtaining a sufficiently high external coupling whilst maintaining the cavity’s quality factor when this coupling is switched out,
represents an engineering challenge --though a surmountable one. The optimal switching time is a happy balance between (i), (ii) and (iii). 

\begin{figure}[htpb]
\includegraphics[width=1.0\columnwidth]{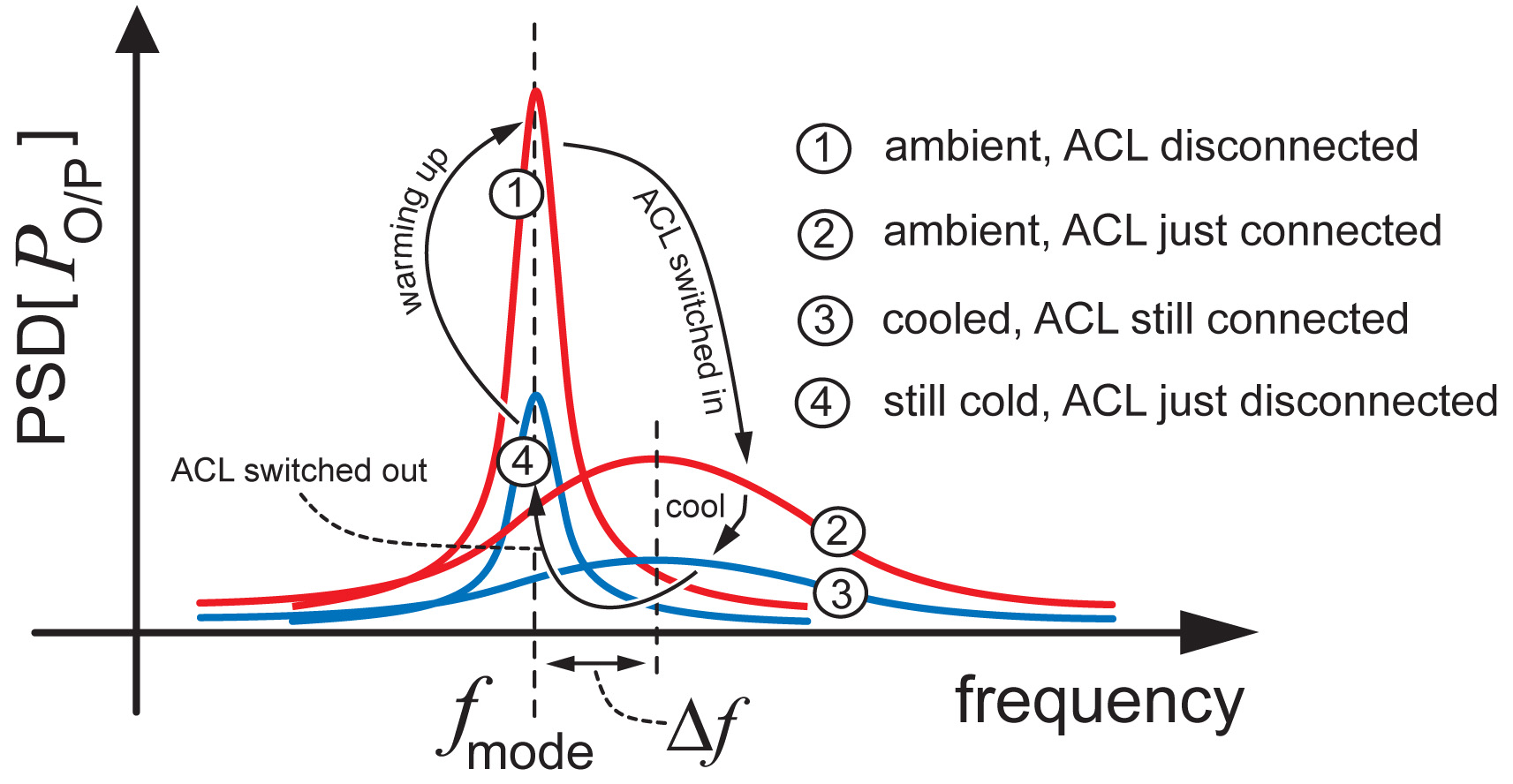}
\caption{\label{fig:frequency_domain}
Active pre-cooling in the frequency domain. The vertical axis is the power spectral density (PSD) of the noise signal extracted by a (small) monitoring probe inserted into the cavity (like what is shown at the bottom right of Fig.~\ref{fig:Siegman_useful}). Here, the linewidth of the thermal noise associated with a mode is inversely proportional to the mode's loaded $Q$. The area under the curve for curves of the same color is the same. }
\end{figure}

The well-known Landau-Zener (L-Z) model\cite{zener1932,glasbrenner2023} provides quantitative insight into the location of this balance. Since, for the relatively high-$Q$ (by room-temperature standards) sapphire-loaded dielectric resonator that we use in our demonstration, the distances in frequency between the operational microwave mode and its neighboring ones correspond to many thousands of cavity linewidths, a switching time that is a tenth or a hundredth of the cavity lifetime easily satisfies adiabaticity according to the L-Z formula (for the probability of a non-adiabatic transition).
Note also that even if adiabaticity is compromised by using a very high over-coupling factor (so perturbing the frequencies and shapes of the cavity electromagnetic modes) and/or a too-rapid switchover, the pre-cooling effect is still protected by the fact that those modes closest in frequency to the operational mode, and thus those that stand to get substantially mixed with it (at anti-crossings),
will also lie within the bandwidth of the ACL, 
and thus stand to be cooled by it also (though admittedly their couplings to the ACL and thus their cooling depths will vary). 
The mixing in of these nearby modes with the operational mode should thus not substantially raise the temperature.
In practice, there is a limit to the amount of switchable external coupling (usually in the form of metal stubs or loops) 
that can be added to a cavity without reducing the intrinsic quality factor (\textit{i.e.}~photon lifetime)  of its modes
when the external coupling is switched out (disconnected). 
The frequency shift \( \Delta f\) associated with the imperfect switching in and out of an overcoupled ACL,
is shown in Fig.~\ref{fig:frequency_domain}.
If the cavity’s mode structure gets excessively scrambled, more distant “hot” modes beyond the ACL’s cooling bandwidth risk getting mixed into the operational mode at switchover so raising its temperature.
We point out here parenthetically that the engineering challenge of achieving a large change in coupling while
incurring only  a small \( \Delta f\)   (or at least the avoidance of anticrossings and mode scrambling) 
draws parallels to the realization of switchable attenuators exhibiting low dispersion, for which proven design strategies/solutions are known. 
Based on our crude L-Z analysis, there would appear to be at least two decades of headroom for raising \( \kappa_\text{over} \)  to the point of “diminishing returns”, where it is no longer the external coupling but the noise temperature of the room-temperature ACL itself that limits the cooling depth. Upon hitting this floor, the only route to lowering the mode temperature still further would be the design and realization of improved low-noise transistor gates/channels, or whatever other active device structure or state of matter (beyond 2-DEGs) that can be co-opted to serve as an improved ACL, whose noise parameters can more severely reduce \( T_{\text{cold}}  \)  as per Eq.~\ref{lnaIPnoise}.

\textit{Further experimental details}:~Our implemented experimental set-up is shown in Fig.~\ref{fig:fig2}. However, when solely demonstrating mode cooling, (i) the cavity need not be loaded with an EPR sample, and (ii), the EPR bridge's circulator can be removed with the monitoring LNA (inside the dashed green box in Fig.~\ref{fig:fig2}) connected directly to the critically-coupled port's stub tuner. Our resonator is a cylindrical brass cavity, internally plated with silver, whose internal volume is partially filled by a concentrically located ring of monocrystalline sapphire (supplied by J-Crystal Photoelectric Technology, China) whose c-axis lies parallel to the cavity's axis of rotational symmetry\cite{wu2020invasive}.
This ring supports a TE\(_{01\delta}\) mode at 1.45~GHz, whose intrinsic quality factor \(Q_\text{cav.}\)~=~164,000,
providing a loaded quality factor at critical coupling (\( \kappa_\text{crit.} = 1 \)) of \(Q_\text{crit.} = Q_\text{cav.} /2\)~=~82,000.
The frequency \(f_0\) and field profile of this mode were
accurately modeled using COMSOL Multiphysics\cite{oxborrow2007traceable}.
Unlike the superheterodyne receiver (incorporating a narrowly filtered 70-MHz IF strip) used in the previous cavity-cooling works by Wu and Ng \textit{et al.}\cite{wu2021bench,ng2021quasi}, we here use a simpler high-gain homodyne receiver to monitor the amplitude/excitation of the TE\(_{01\delta}\), recording the receiver's output into a Tektronix TBS 1102B-EDU oscilloscope.

\begin{figure}[b]
\includegraphics[width=\columnwidth]{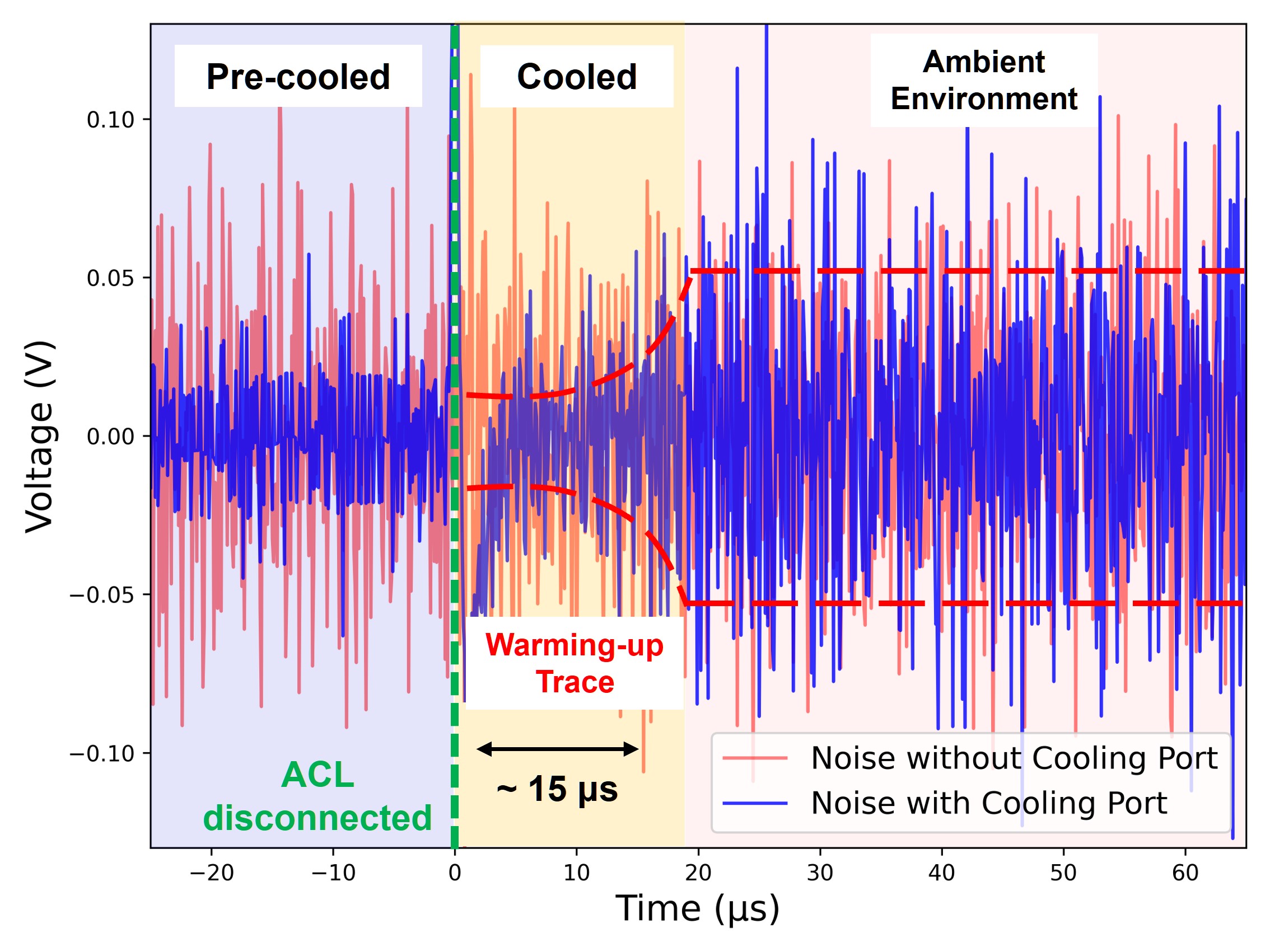}
\caption{\label{fig:fig4} 
Increase in measured noise over time as the experimental cavity's monitored microwave mode warms up from ``pre-cooled'' to ambient conditions. The two red dashed lines track the envelope of the voltage noise.
Upon curve-fitting to this envelope (see Fig.~S9 in the Supplemental Material),
the characteristic time of the warming up is determined to
be \(\tau_{\text{warm up}}^{\text{exp.}} \approx 9 \pm 1 \mu\)s.
} 
\end{figure}

It is worth pointing out that both the cooling and the monitoring of the TE\(_{01\delta}\) mode \textit{could} alternatively be implemented using just a single port attached to a single LNA, \textit{if} this port's coupling to the mode \textit{were} made dynamically adjustable. The LNA would be alternatively connected (``toggled''),
at sufficient speed and glitch-free, between (say) a larger and a smaller coupling loop. However, given that our cavity already had two separate coaxial ports (feedthroughs), it was simpler,
with the components immediately available to us, to use both of these ports set
to different fixed couplings \( \kappa \), namely:
(i)~an over-coupled ``cooling'' port (\(\kappa = \kappa_{\text{over}} =3.8\)), connected through a low-insertion-loss single-pole double-throw (SPDT) switch (\textit{viz.}~a Qorvo RFSW1012 mounted on an evaluation board) to either the ACL (\textit{viz.}~the input of a Qorvo QPL9547 LNA mounted on an eval.~board) or to an ``open'' termination --the choice being determined by the control voltage applied to the RFSW1012 switch; and (ii)~a ``monitoring/interrogation'' port, critically coupled ($\kappa = \kappa_\text{crit.} = 1$) to the TE\(_{01 \delta}\) mode through a Maury Microwave 1819A triple-stub tuner, and used to monitor the mode's field amplitude (and thus noise temperature). These two different ports are depicted, conceptually, in Fig.~\ref{fig:fig1}(b); their experimental embodiments are displayed in Figs.~\ref{fig:fig2}(a) and \ref{fig:fig2}(c).
The coupling factors of both ports (cooling and monitoring), as well as the
quality factor of the sapphire resonator's  TE\(_{01 \delta}\) mode, were accurately determined using an HP8753C vector
network analyzer (VNA)\cite{kajfez1984q, chua2003accurate}.

\begin{figure}[htpb]
\includegraphics[width=1\columnwidth]{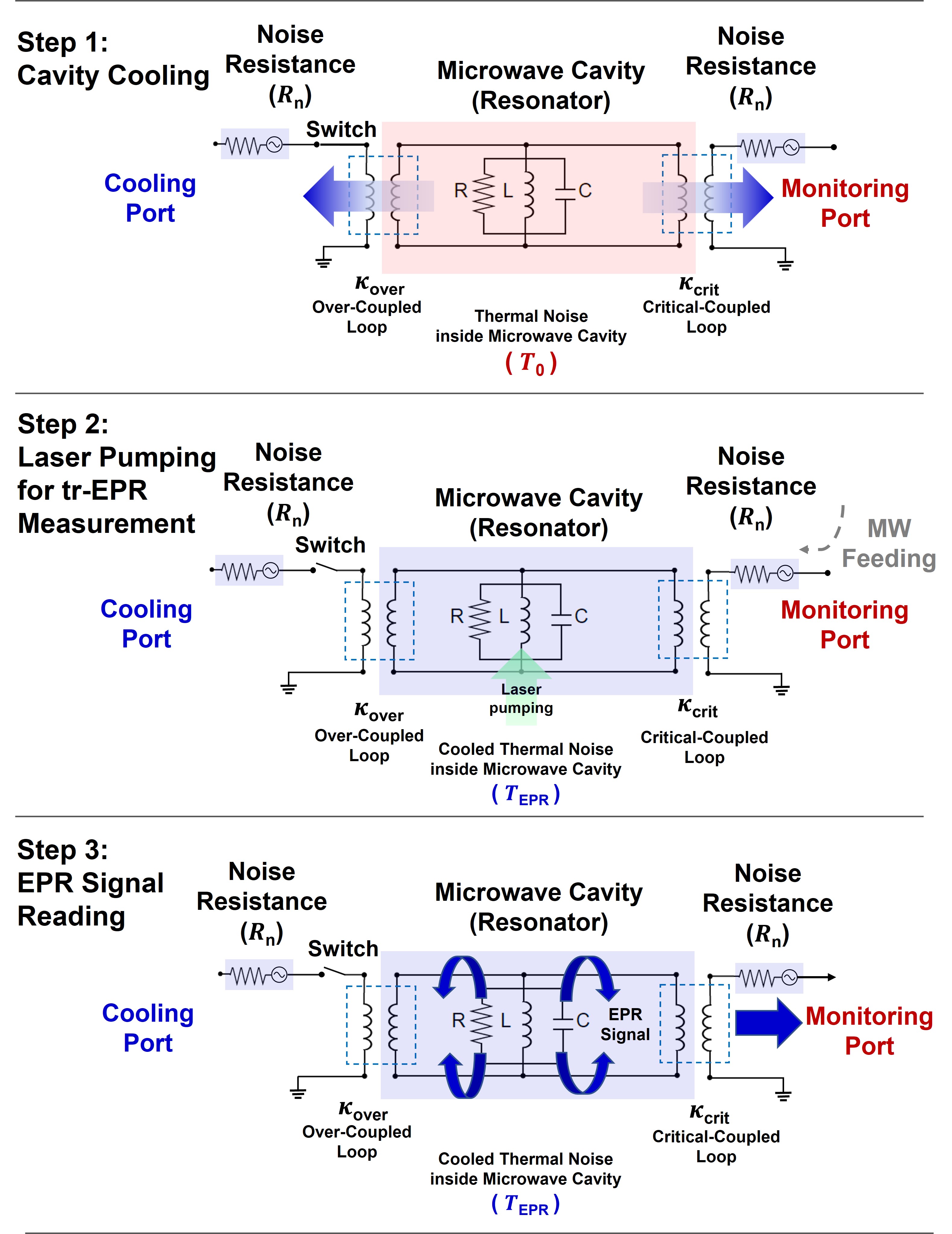}% 
\caption{\label{fig:figStep123}
Active pre-cooling as applied to time-resolved (tr-) EPR: In Step 1, an LNA connected to an over-coupled cooling port extracts thermal noise (energy) from the microwave cavity (represented here as an LCR circuit); this also pulls the cavity mode's resonance frequency away from \(f_0\) --slightly. Next (Step 2), a switch disconnects the cooling LNA, bringing the mode back on resonance (\textit{i.e.}, its frequency coincident with the EPR transition's line center at 1.45 GHz). Instantaneously with this microwave switching, a q-switched optical pulse (around 7 nanoseconds in duration) of 532-nm pump light is shone on the EPR sample within the cavity. Step 3: a continuous-wave (CW) interrogating microwave tone is thereupon applied via a circulator to the cavity's monitoring port, as shown in Fig.~\ref{fig:fig2} (c), which, upon re-emerging from the cavity through the circulator's third arm, provides a tr-EPR signal characterizing the EPR sample.}
\end{figure}

It is worth pointing out how the monitoring LNA itself helps to cool the mode that it monitors.  This would not happen if, as is often done, a microwave isolator were inserted between the monitoring LNA and the cavity --so as to ``protect'' the cavity
from external noise. Noise (and thus sensitivity) wise, it is always advantageous to connect the monitoring LNA directly to the cavity, so long as its input noise temperature is lower that the physical temperature of the isolator's internal termination.  

 \begin{figure*}[htpb]
\includegraphics[width=2\columnwidth]{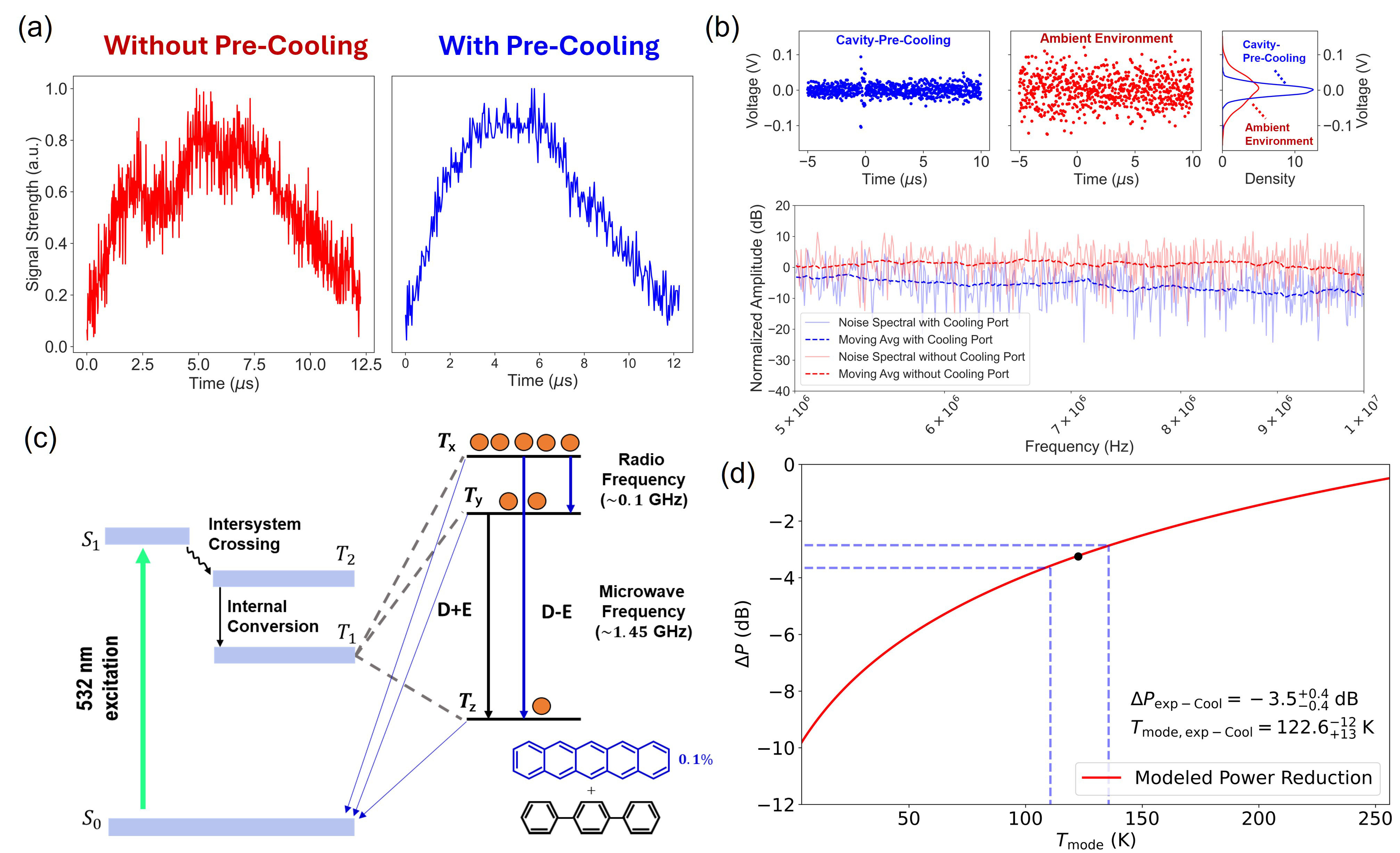}
\caption{\label{fig:fig6}(a)~Zero-field (ZF) tr-EPR signal of Pc:PTP's \(T_\text{x}-T_\text{z}\) transition\cite{wu2019unraveling} without (red) and with (blue) pre-cooling.
(b)~Thermal noise (in the time domain) extracted from each of these ZF tr-EPR signals and, below, their corresponding spectral densities; as before, blue and red correspond to the pre-cooled (with) and ambient (without) conditions, respectively. 
(c)~Simplified Jablonski (= energy-level) diagram of the ZF sublevels (\(T_\text{x}\), \(T_\text{y}\), \(T_\text{z}\)) of pentacene's triplet ground state, where D and E denote the ZF splitting (ZFS) parameters.
(d)~Modeled noise power reduction
(\(\Delta P\)) at the homodyne receiver's input as a function of the noise temperature of the EPR cavity's operational mode, \(T_{\text{mode}}\). Since its uncertainty is lower, the estimated depth of cooling presented here (\(\Delta P_\text{exp-Cool}\)) corresponds to that achieved by
active pre-cooling alone (see Table~S2 in the Supplemental Material\cite{supplemental}).
The depth's standard error is indicated by dashed cornflower-blue lines.}
\end{figure*}

The reduction in thermal noise will persist so long as the cooling LNA is connected. Upon disconnecting it and connecting the over-coupling loop (port) instead to a reflective ``open'', the mode regains its higher quality factor, namely 
\(Q_\text{crit.}\),
and warms up on a timescale set by the mode's photon lifetime\cite{aspelmeyer2014cavity},
namely \( \tau_\text{warm up} = \frac{Q_\text{crit.}}{2 \pi f_0}. \) Note that this is substantially
longer than the time required for cool down. [Parenthetically: any perfectly reflecting impedance, like
a ``short'', could serve just at well as the open that we employ here.]
Since the monitoring port is critically coupled to the TE\(_{01\delta}\) mode,
this mode's temperature will settle at the (equally weighted) average of $T_{\text{hot}} = 290$~K and $T_{\text{mon.}}$,
namely \( T_{\text{mode}}^{\text{ambient}} = 257.6\)~K for our particular experiment.

\section{Results}

\textit{Preliminaries}: We verified the noise characteristics of our cooling LNA (\textit{viz.}~a Qorvo QPL9547EVB-01 operating at room temperature) using the industry-standard \mbox{Y-factor} method\cite{agilentAN57-1}.
This included connecting the~LNA's input to a passive load, well matched to 50~\( \Omega\),
both at room temperature and when immersed in a small dewar filled with liquid nitrogen. 
Within experimental error (taking noise generated by losses along lengths of line
into account), our own measurements were wholly consistent with the manufacturer's specification sheet. 
When used as an amplifier, this LNA exhibits, at 1.45~GHz, an impressive noise figure of just 0.182 dB
(for \( \Gamma_\text{s} = 0 \); referenced on the device's leads),
equivalent to an effective noise temperature (referred to input) of just \( T_\text{eff.} = 12.4\)~K;
see Table~S1 in the Supplemental Material \cite{supplemental} for this LNA's full noise parameters
interpolated to 1.45~GHz, and Refs.~\cite{dunleavy2009,modelithicsAN-60-040} for associated definitions.
Note that \( T_\text{eff.} \), calculated using Eq.~\ref{lnaOPeffnoisetemp}, is
substantially smaller that the temperature of the same device when configured as a cold load,
namely \(T_{\text{cold}}\) \( =\) 29.1~K, calculated using Eq.~\ref{lnaIPnoise}.
Since the QPL9547 MMIC was designed/optimized by its manufacturer
for use as a LNA, this should not be hugely surprising. 

Experimentally, we observed a reproducible transient on the EPR (homodyne) signal as the SPDT switch (\textit{viz.}~a Qorvo RFSW1012EVB-01) opened and disconnected the pre-cooling LNA from the microwave cavity.
This artefact, lasting less than 3~\(\mu\)s in our experiment,
could be substantially removed by measuring it for multiple switch openings then subtracting the average from our measured traces.

\textit{Mode Cooling}:
The effect of connecting and then disconnecting the over-coupled cooling LNA to
our cavity's TE\(_{01\delta}\) mode at \(f_0 =\)~1.45~GHz is shown in Fig.~\ref{fig:fig4}.
From the difference in the time average of the measured noise voltage squared for the mode
under cooled and ambient conditions, the depth of cooling is estimated to be \(\Delta P_{\text{exp-Cool}}= -3.5 \pm 0.4\)~dB.
This agrees reasonably well with both our theoretical estimate
\(\Delta P_{\text{model}}\) and our analysis using the noise spectrum from experimental time-resolved-EPR data, namely \(\Delta P_{\text{exp-EPR}}\),
as shown in Fig.~\ref{fig:fig6}(b).
All our estimates are summarized in Table~S2 of the Supplemental Material\cite{supplemental}.
Upon curve-fitting to the rise in the measured noise voltage squared as a function of time,
the characteristic time quantifying the persistence of the mode's coldness after
disconnection from the cooling LNA is estimated to be 
\(\tau_{\text{warm up}}^{\text{exp.}} = 9 \pm 1~\mu\)s; 
see Fig.~S9 in the Supplemental Material\cite{supplemental}.

\textit{Applicaton to EPR}: The protocol used to implement active pre-cooling (APC) is depicted in Fig.~\ref{fig:figStep123} and consists of three stages: initial noise reduction (mode cooling) with the over-coupled port connected to an ACL.
This is followed by disconnection of the ACL,
re-tuning the operational mode of the cavity back onto the
line \mbox{center} of the target EPR transition.
A nano-second optical pump pulse [provided by an optical \mbox{parametric} \mbox{oscillator} (OPO)
pumped by a q-switched laser]
is then \mbox{immediately} applied, followed by a continuous-wave (CW) microwave
tone, whose reflection off the now critically-coupled
cavity provides, upon homodyne detection (with filtering and amplification both before and after downshifting), a time-resolved EPR signal.

\begin{figure}[htpb]
\includegraphics[width=1\columnwidth]{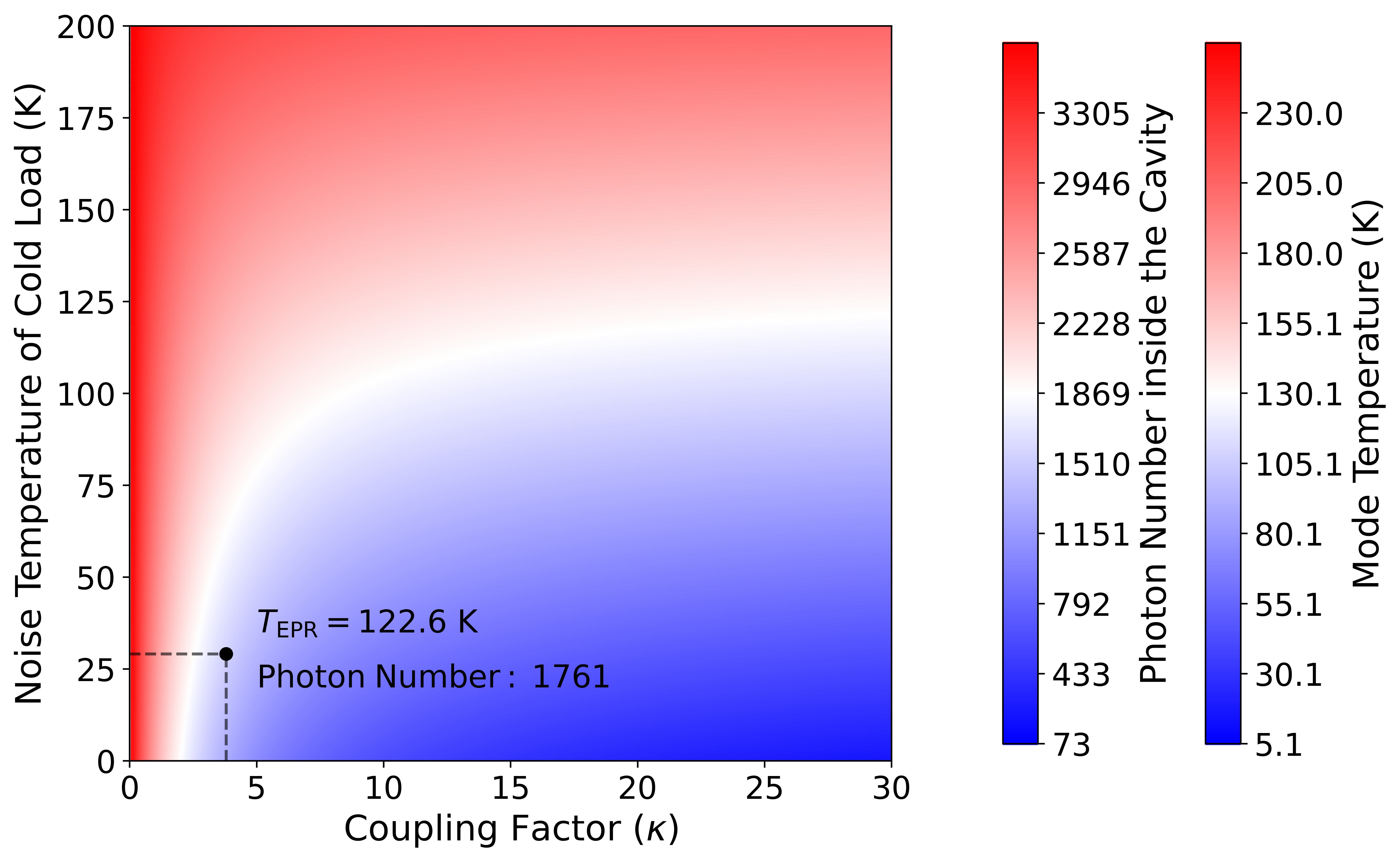}
\caption{\label{fig:fig3}
Modeled average number of thermal photons occupying the cavity's  TE\(_{01\delta}\) mode and the mode's corresponding temperature (\(T_\text{mode}\)) as a function of the cold load's own noise temperature (\(T_{\text{cold}}\)) and its coupling factor (\(\kappa\)) to the mode.
In our experimental demonstration, this mode's temperature is reduced from 257.6~K to 122.6~K (measured),
the latter temperature corresponding to an average occupancy of 1761 photons.}
\end{figure}

To demonstrate the utility of active pre-cooling, we studied the signal-to-noise ratio (SNR) and noise spectrum of traces obtained from performing tr-EPR on a photo-excited sample of pentacene-doped \textit{para}-terphenyl (Pc:PTP) at room temperature in zero applied magnetic field. This sample and the complete spectrometer around it are shown in Fig.~\ref{fig:fig2}. Fig.~\ref{fig:fig6}(a) displays tr-EPR signals for the Pc:PTP triplet's Tx-Tz transition
at 1.4495~GHz 
[as shown in Fig.~\ref{fig:fig6}(c)] both without (red) and with (blue) pre-cooling. 
The noise on each of the measured tr-EPR signals shown in Fig.~\ref{fig:fig6}(a) was extracted\cite{hassan2010reducing} from the causal EPR response by first (i)~generating a smoothed-out version of the measured data by boxcar averaging, the width of each averaging box being set at 100~ns, then (ii)~subtracting this version from the directly measured data to provide the traces shown at the top of Fig.~\ref{fig:fig6}(b). Their respective spectral densities (modulus squared of Fourier transform) are displayed underneath. Here, only the section dominated by white noise, above the corner frequency of the
observed \(1/f\) noise (at around 1 MHz) generated by the receiver's LNAs, is shown. These plots attest to the  significant reduction in white (thermal) noise that pre-cooling can provide, the SNR (immediately after pre-cooling) being enhanced  by \(3.5^{+0.4}_{-0.4}\)~dB.
The relationship between the reduction in the noise power and the achieved noise temperature of the TE\(_{01\delta}\) mode at 1.45~GHz is shown in Fig.~\ref{fig:fig6}(d). In our demonstration, the mode temperature drops
from an initial \(T^{\text{ambient}}_{\text{mode}}~=~257.6\)~K
to a pre-cooled \(T_{\text{mode}}^{\text{cooled}}~=~122.6\)~K (measured), corresponding to a reduction in the expected number of thermal photons from 3699 to 1761.

The plot shown in Fig.~\ref{fig:fig3} indicates that, at room temperature, the number of thermal photons in a mode at \(1.45 \, \text{GHz}\) can be reduced to just a few hundred (\textit{cf.}~ref.~[\onlinecite{raimond1982collective}]) given a sufficiently over-coupled port (\(\kappa_{\text{over}}~>~10\)) and quiet LNA ( \(T_{\text{cold}} < 20 \, \text{K}\)). 

Restricting ourselves to work done on bench-tops at room temperature, we here briefly compare the performance
achieved by our demonstration of active pre-cooling (APC) to that achieved by removing thermal photons
via stimulated absorption in an absorptively spin-polarised medium resident
inside the cavity.
In the literature,
this latter approach has been dubbed ``masar''\cite{wu2021bench} or ``anti-maser''\cite{blank23}. 
Using photo-excited Pc:PTP as the absorber, Wu \textit{et al.}\cite{wu2021bench} cooled a TE\(_{01\delta}\) mode at 1.45 GHz down to around 50~K --though only in bursts. Using NV\(^-\) diamond, Ng~\textit{et al.}\cite{ng2021quasi} and Day~\textit{et al.}\cite{day24} cooled the same type of mode continuously to 188~K at 2.9~GHz and to 70~K at 9.8~GHz, respectively. Given that these temperatures straddle what we here achieve using APC, one may conclude that, currently, there is no clear winner. Both approaches can assuredly be improved upon through focused material science and device engineering.  
Beyond cost and convenience, a salient advantage of APC is that the disturbance on the microwave cavity imposed by the cooling mechanism can be removed at the flick of a (microwave) switch. Methods based on internal stimulated absorption are, in contrast, beholden to the particular spin dynamics of the cooling medium. One could however use an \textit{external} masar/anti-maser
(operated either in CW mode or quasi-continuously\cite{wu2020quasi}) as the ACL to implement pre-cooling instead of a HEMT-based LNA,
with the connection between the resonator and the anti-maser's coupling port gated by a microwave switch.
As with other pulse-based techniques at microwave frequencies, a successful implementation of APC relies heavily on the availability of a sufficiently fast, low-loss and glitch-free switch.  

To conclude, we have demonstrated a method for actively pre-cooling the modes of a microwave cavity by temporarily over-coupling them to the input of a low-noise amplifier (LNA) serving as an active cold load (ACL). Beyond tr-EPR, other pulsed-EPR techniques such as electron spin-echo envelope modulation (ESEEM), hyperfine sublevel correlation spectroscopy (HYSCORE), and pulsed electron-electron double resonance (PELDOR) could be accelerated by APC, reducing measurement times\cite{poole2019electron}. 
Furthermore, our analysis identifies how an LNA that is \textit{directly} coupled to a cavity (\textit{i.e.},~when no ``protective'' isolator/circulator is inserted in between), can play an active and beneficial role in reducing the number of thermal photons occupying the cavity's modes (within the LNA's bandwidth), so reducing measurement noise.

The use of \textit{cryogenic} ACLs, operating at (say) liquid-helium temperature yet providing input noise temperatures and thus mode cooling down into the mK regime, is perfectly conceivable. Unfortunately, due to self-heating within the device's active channel/layer, the noise temperatures of what are currently the quietest semiconductor-based LNAs hit, upon physical cooling, plateaux of a few K\cite{ardizzi22,ardizziPHD2022,zeng24}. The construction of effective
cryo-ACLs to achieve mK-level mode cooling may well require new materials and/or architectures (and the equipment to fabricate such), optimized expressly for mode-cooling --as opposed to the repurposing of existing devices (designed with other applications mind) like what  our current exploration has been based upon.
At a more general, sociologically level, acknowledging that the levels of cooling we have here achieved using APC are still far away from the single-photon regime, it is hoped that this paper induces greater
cross-fertilization between wisdom embedded within the discipline of microwave radiometry\cite{cakaj2011antenna,weissbrodt2017}, activities across the EPR/NMR communities\cite{hyde2005trends, chen2024unlocking}, and hardware innovations in quantum computing/detection, particularly those targeting room-temperature operation\cite{henschel2010cavity,pezzagna2021quantum}. 

Future work will focus on the engineering of cavity designs that allow for higher over-coupling factors, whilst providing sufficient adiabaticity and switch-over speed, such that the cooling depth is limited more by the noise temperatures of available ACLs. Integrating these advancements with existing microwave-based quantum sensing instruments beyond time-resolved EPR would demonstrate the broader utility of our approach. 

Acknowledgments: The authors would like to thank Wern Ng, now at UC Berkeley, and Hao Wu, at the Beijing Institute of Technology (BIT), for helpful discussions. This work was supported by the U.K. Engineering and Physical Sciences Research Council through Grants Nos. EP/K037390/1 and EP/M020398/1. K.C. acknowledges financial support from the Taiwanese Government Scholarship to Study Abroad (GSSA). 

%\nocite{*}

\bibliography{PRA_CPC}

%\documentclass[%
% reprint,
% superscriptaddress,
%groupedaddress,
%unsortedaddress,
%runinaddress,
%frontmatterverbose, 
%preprint,
%preprintnumbers,
%nofootinbib,
%nobibnotes,
%bibnotes,
% amsmath,amssymb,
% aps,
%pra,
%prb,
%rmp,
%prstab,
%prstper,
%floatfix,
%bookmarks=true, pdfnewwindow=true, colorlinks=true, linkcolor=xlinkcolor, citecolor=xlinkcolor, filecolor=xlinkcolor, urlcolor=xlinkcolor, final=true,
%]{aastex631}

%\documentclass[%
% reprint,
% superscriptaddress,
% amsmath,amssymb,
% aps,
% pra,prb,rmp,prstab,prstper,
% floatfix,
% bookmarks=true, pdfnewwindow=true, %colorlinks=true, linkcolor=xlinkcolor, %citecolor=xlinkcolor, filecolor=xlinkcolor, %urlcolor=xlinkcolor, final=true,
%]{aastex631}

%\usepackage{graphicx}% Include figure files
%\usepackage{dcolumn}% Align table columns on decimal point
%\usepackage{bm}% bold math
%\usepackage{hyperref}% add hypertext capabilities
%\usepackage[mathlines]{lineno}% Enable numbering of text and display math
%\linenumbers\relax % Commence numbering lines
%\usepackage{booktabs}
%\newcommand{\vdag}{(v)^\dagger}
%\newcommand\aastex{AAS\TeX}
%\newcommand\latex{La\TeX}
%\newcommand{\tablesuffix}{S}
%\graphicspath{{./}{figures/}}

%\begin{document}

%\title{Supplementary Material:\\ 
%Overcoming the Thermal-Noise Limit of Microwave Measurements\\
%by Pre-cooling with an Active Cold Load}

%\author[0000-0002-6575-7034]{Kuan-Cheng Chen}
%\affiliation{Department of Materials, Imperial College London\\
%Exhibition Road, London SW7 2AZ, United Kingdom}

%\author{Mark Oxborrow}
%\email{m.oxborrow@imperial.ac.uk}

\clearpage
\onecolumngrid

%\appendix*
\setcounter{page}{1}
\renewcommand\thepage{s\arabic{page}} 

\setcounter{section}{0}

\renewcommand\thefigure{S\arabic{figure}} 
\setcounter{figure}{0} 

\renewcommand\thetable{S\arabic{table}} 
\setcounter{table}{0}

\textbf{\hspace{5cm}  \large{SUPPLEMENTAL MATERIAL} }

\textbf{\section{EXPERIMENTAL METHOD - DETAILS} }\label{sec:exp}

\begin{figure}[!b]
\centering
%\figuretag{(S1)}
%\tag{(S1)}
\includegraphics[width=1\columnwidth]{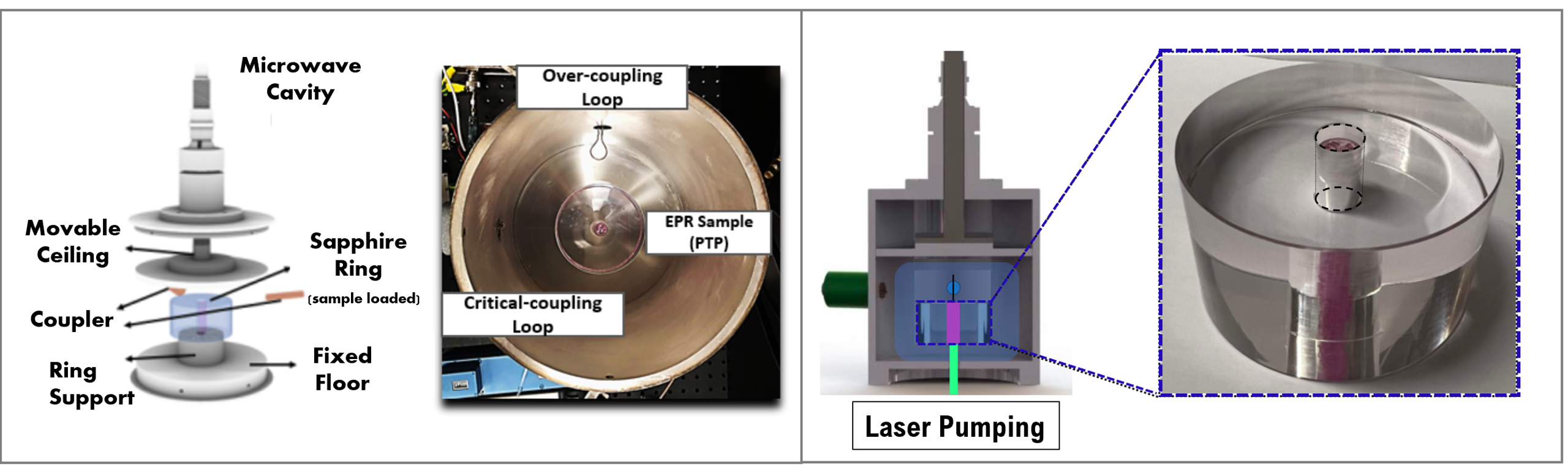}
\caption{\label{FS1} The left image presents the 3D CAD assembly of the microwave cavity (without microwave walls), featuring a movable ``ceiling'' (= plunger), raised and lowered by a rotating a helical screw (to which the ceiling is mechanically attached) seated in a matching helical thread tapped into the cavity's ``roof'' along the cavity's cylindrical axis. The photo to the right of the drawing shows the cavity's innards with the ceiling and roof removed. The cavity includes two ports: an over-coupling loop (at 12 o'clock in the photo) and a smaller critical-coupling loop (at 9 o'clock) just poking out of the cavity's internally-silver-plated wall. The right figure displays the microwave cavity in 2D cross-section. Here, the high-Q sapphire ring is located concentrically on top of a cross-linked polystyrene stand (not shown); its on-axis bore hole, 10~mm in diameter, is filled with the EPR sample (a 0.1\% PC:PTP crystal).}
\end{figure}

The anatomy of the microwave cavity used in our zero-field time-resolved (tr-) electron paramagnetic resonance (EPR) spectrometer for cavity pre-cooling demonstrations is shown in Fig.~\ref{FS1}. This cavity supports a \( \textrm{TE}_{01 \delta} \) mode at 1.4495 GHz, exhibiting a loaded quality factor of \(Q_\text{loaded} = 82,000\) when critically coupled, corresponding to an unloaded quality factor of \(Q_0 = 164,000\). The cavity consists of five main components: (i) a cylindrical brass enclosure, whose internal walls are plated with silver, (ii) a monocrystalline sapphire ring (supplied by J-Crystal Photoelectric Technology, China) exhibiting low dielectric loss, (iii) an EPR sample, namely a 0.1\% pentacene-doped para-terphenyl (PC: PTP) crystal, located in the ring’s bore, (iv) an over-coupling loop (coupling factor = 3.8)
connected through a single-pole-double-throw (SPDT) RF switch [\textit{viz.}~a Qorvo RFSW1012PCK-411 mounted on an evaluation (eval.) printed circuit board] to a low-noise amplifier (LNA) [\textit{viz.}~a Qorvo QPL9547EVB-01 on an eval.~board], functioning as an active cold load,
and (v) a critical-coupling loop connected, via a stub tuner, to a second identical LNA (another QPL9547EVB-01), functioning as the front-end pre-amplifier of a tuned homodyne receiver. The sample receives a Q-switched optical pulse at 532~nm, approximately 5.5 ns in duration and 2 mJ in energy, at a 10 Hz repetition rate, from an integrated Nd:YAG-pumped type-II-phased-matched $\beta$-barium-borate (BBO) Optical Parametric Oscillator (OPO) (\textit{viz.}~a Litron Aurora II Integra).
The cavity's internal cylindrical sapphire ring measured 40 mm in height with outer and inner diameters of 68 mm and 10 mm, respectively. The optical ``c''-axis of the sapphire (a monocrystal) was aligned with the cylindrical axis. 
By rotating a screw, the cavity's axial length (\textit{i.e.}~the internal height of the cavity's ceiling above its ``floor'') could be varied from 150 mm down to 95 mm, allowing wide adjustment in the frequency of the cavity's \( \textrm{TE}_{01 \delta} \) mode.
The sapphire ring's hollow supporting pillar was made of cross-linked polystyrene,
a solid material offering both low dielectric loss and
(unlike polytetrafluoroethylene or polyethylene) precision-machineability. 
This pillar was securely seated into a central hole in the cavity's floor, holding the sapphire ring 16 mm above the floor. The pillar's bore allowed for the transfer of pump light, up the cavity's axis, to the EPR sample. 
The cavity's silvered internal surfaces were finely buffed with an abrasive pad, removing patina, then solvent-wiped clean, to leave a gleaming finish.

By constructing a suitable model in COMSOL Multiphysics (a finite-element-based PDE solver) \cite{oxborrow2007traceable},
the magnetic mode volume \(V_{\text{mode}}\) of the microwave cavity's \(\text{T}_{01 \delta}\) mode was calculated to be 29~cm\(^3\). Experimentally, using a vector network analyser
(= VNA, \textit{viz.}~an HP-8753C), the cavity's loaded quality factor (\(Q_\text{loaded} = 82,000\)) and the coupling factors of each of its two ports (\(\kappa_{\text{over}}=3.8\) and \(\kappa_{\text{crit.}} = 1\)) were inferred by the method(s) of Kajfez and Hwan \cite{kajfez1984q}. The PC:PTP crystal, used as our tr-EPR test sample, was grown by the vertical Bridgman method; it  fitted into the 10-mm bore of the sapphire ring. The pulsed optical beam from a Q-switched OPO was directed into the microwave cavity via a multimode optical fiber, entering through an 8-mm coaxial hole in the centre of the cavity's floor. 

The homodyne receiver for our tr-EPR measurements used a setup akin to Wu et al.'s previous work \cite{wu2019unraveling}, albeit with the microwave cavity (resonator) now connected to an additional port (for cooling) and with additional routing switches and associated synchronization circuitry -see Fig.~4(c) in the main paper and Fig.~\ref{FS2} below.
A sync.~pulse from the OPO's Q-switch driver is used to
trigger two pulse generators [both Thandar TG105, one triggered by (= ``slaved to'') the other] providing appropriate TTL voltage transitions  (``edges'') to connect/disconnect the precooler and switch on/off the EPR spectrometer's interrogating microwave tone. The most critical switch, needed to perform a clean, abrupt disconnect of the precooling LNA, was a Qorvo RFSW101 SPDT switch, chosen for its low insertion loss, high isolation and sufficient speed; this routes the cooling port's over-coupling loop first to the active cold load (embodied as the input of a 
Qorvo QPL9547 LNA) for cooling and, subsequently, to an ``open'' load (reflecting power back into the cavity, effectively neutralizing the port) for the tr-EPR measurement, when the interrogating microwave tone is applied through the cavity's critically-coupled monitoring port.  To achieve sufficient measurement sensitivity, the homodyne receiver employs a cascade of legacy r.f.~amplifiers (total gain \(\approx 70\) dB). Upon mixing (= homodyning) down to baseband, the transient tr-EPR signal is channeled (via a DC block) into a fast DC amplifier (\textit{viz.}~a~Comlinear E103-I-BNC). All components operate under ambient lab conditions.
\begin{figure}[htbp]
\centering
% \figuretag{(S2)}
\includegraphics[width=1\columnwidth]{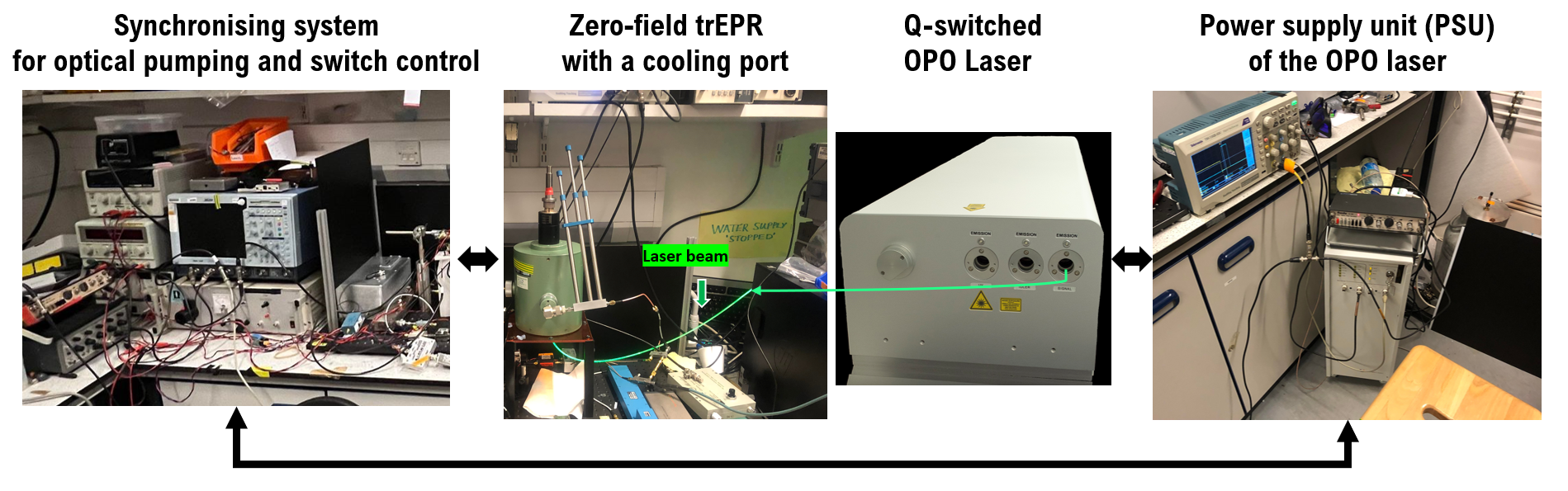}% Here is how to import EPS art
\caption{ \label{FS2} The experimental rig (here emphasizing its electrical synchronization and light path) encompassing, from right to left: the OPO's power supply unit (PSU), the OPO laser head itself, the resonator and it surrounding microwave ``plumbing", and the tr-EPR spectrometer's support electronics. }
\end{figure}

\clearpage

\begin{figure}[htbp]
\centering
% \figuretag{(S3)}
\includegraphics[width=0.7\columnwidth]{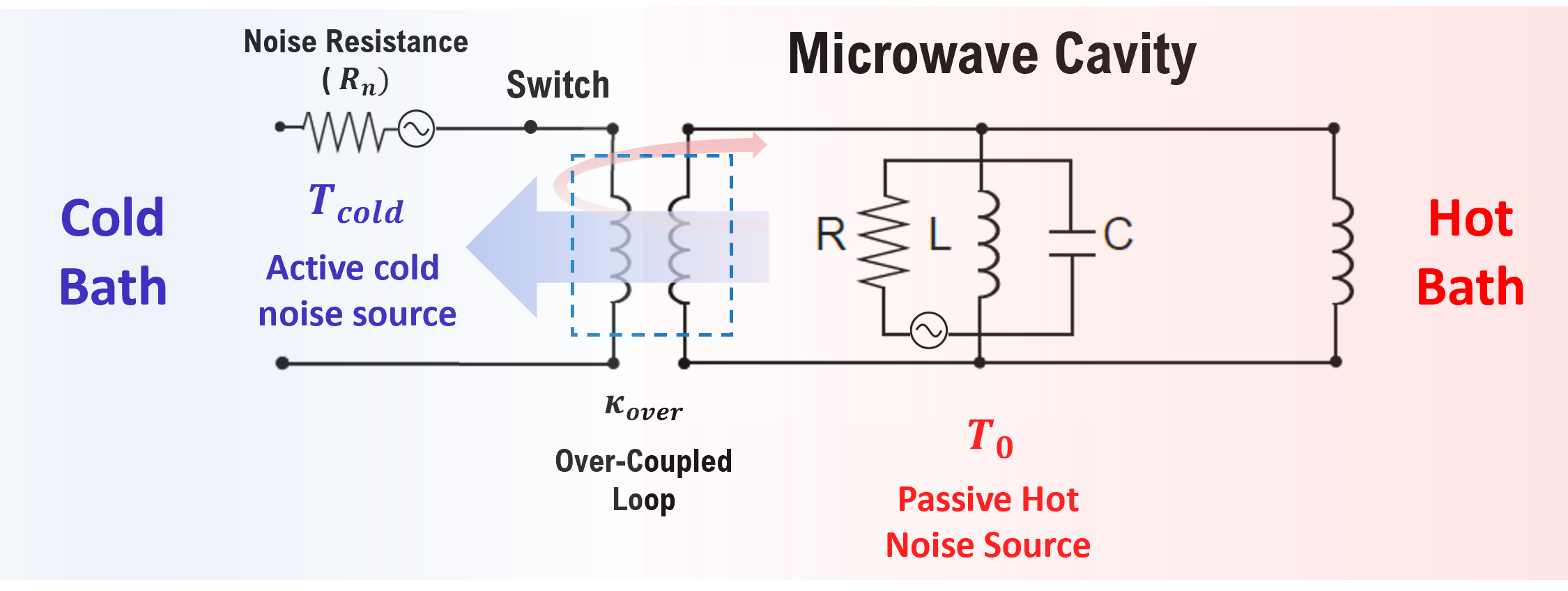}% Here is how to import EPS art
\caption{ \label{FS3}The active pre-cooling technique involves electromagnetically coupling a resonator, that is physically heat-sunk to a room-temperature environment (here referred to as the ``hot bath''), to an active cold noise source or ``load'', here functioning as a ``cold bath''.}
\end{figure}

\textbf{\section{ CONCEPT AND MODEL OF THE ACTIVE-PRE-COOLING TECHNIQUE}} \label{sec:concept}
\textbf{\subsection{ Number of photons inside a microwave cavity}}
Obeying Bose-Einstein statistics, the average number of photons, $\bar{n}$, occupying an electromagnetic mode of frequency \(f_{\text{mode}} = \omega_{\text{mode}} /(2 \pi)\) inside an isolated (zero ports) cavity maintained at a physical temperature of \( T_0 \), is given by $\bar{n} = \left[\exp\left(\frac{h f_{\text{mode}}}{k_\text{B} T_0}\right)-1\right]^{-1}$, where \(k_\text{B}\) and \(h\) are Boltzmann's and Planck's constants, respectively. 
Semiclassically, for a microwave cavity with a single port coupling it to an external cold load, as depicted schematically in Fig.~\ref{FS3}, the number of photons \(q\) occupying the mode
as a function of time \(t\) will obey
\begin{equation} \tag{S1} \label{EqS2}
   \frac{\text{d}q}{\text{d} t} \equiv \dot{q}  = \underbrace{- \omega_{\text{mode}} Q_0^{-1} (q - \epsilon T_0)}_{\colorbox{pink}{interaction with intrinsic ``hot'' heat bath}} -  \underbrace{\omega_{\text{mode}} \kappa Q_0^{-1} (q - \epsilon T_{\text{cold}}) }_{\colorbox{cyan}{interaction with external ``cold'' heat bath}},
\end{equation}
where the value of the constant \(\epsilon = \frac{k_\text{B}}{hf_{\text{mode}}}\) respects Maxwell-Boltzmann equipartition of energy\cite{wu2021bench}.
\(T_{\text{cold}}\) is the noise temperature of the active cold load  used in the experiment;
\(Q_0\) and \(Q_{\text{loaded}}\) are the intrinsic and loaded quality factors of the microwave cavity, where \(Q_{\text{loaded}} = \frac{Q_0}{1+\kappa}\); \(\kappa\) is the port's coupling factor.
This formula can  be trivially extended to include additional ports coupling the microwave mode at different couplings strengths \(\kappa_i\) to additional loads at different temperatures \(T_i\): 
\begin{equation} \tag{S2} \label{EqS2mo}
   \dot{q} = -  \underbrace{\omega_{\text{mode}} Q_0^{-1} (q - \epsilon T_0)}_{\colorbox{pink}{intrinsic bath}}
   -  \sum_{i = 1} \underbrace{\omega_{\text{mode}} \kappa_i Q_0^{-1} (q - \epsilon T_i) }_{\colorbox{cyan}{external bath(s)}}.
\end{equation}

\clearpage

\textbf{\subsection{ Mode temperature inside the cavity}}

\begin{figure}[htbp]
\centering
%\figuretag{(S4)}
\includegraphics[width=0.6\columnwidth]{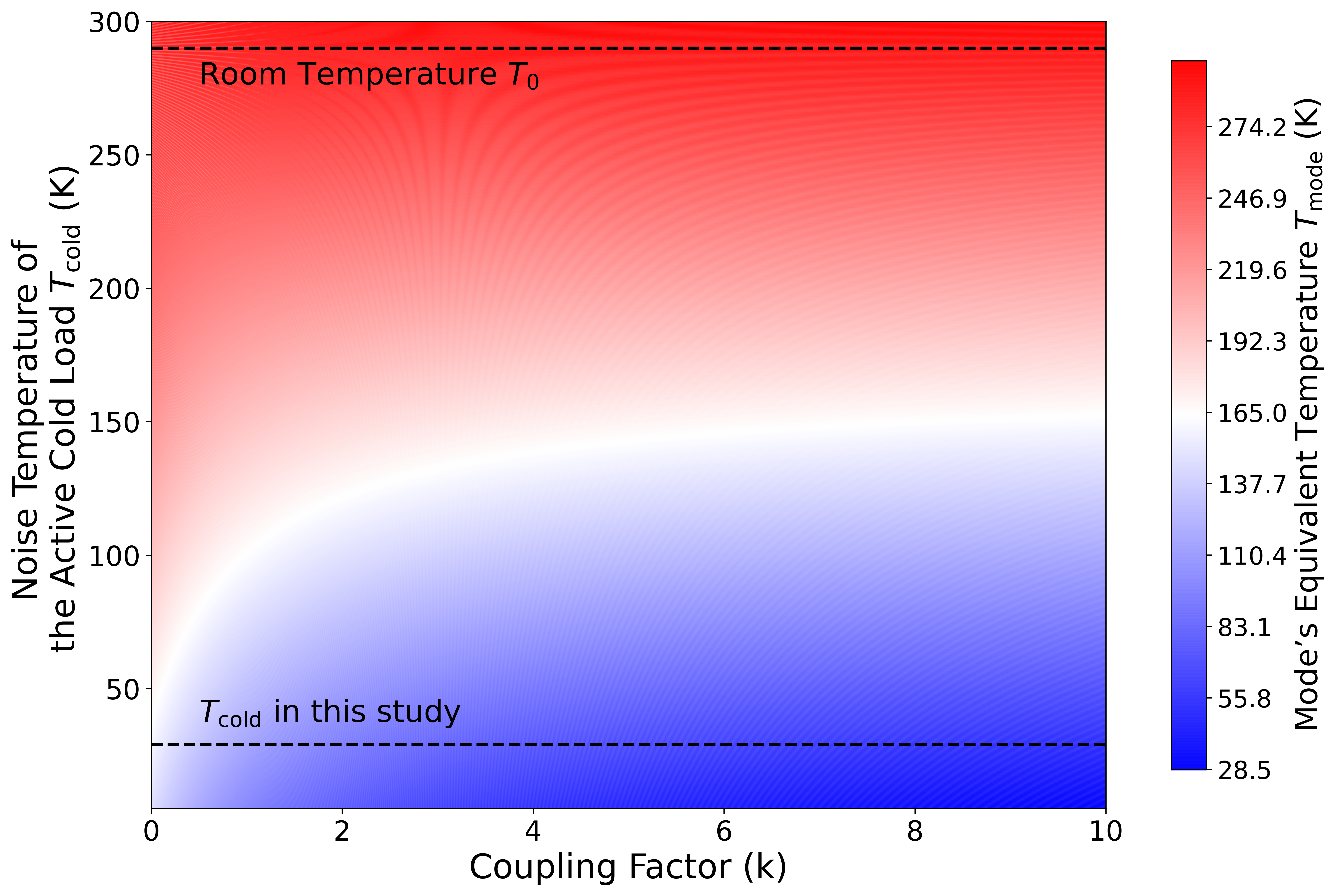}% Here is how to import EPS art
\caption{\label{FS4} 
Modeled equivalent temperature of an electromagnetic mode inside the cavity as a function of the coupling factor \(\kappa\) (horizontal coordinate) and the noise temperature of the cold load \(T_{\text{cold}}\) coupled to this mode (vertical coordinate). }
\end{figure}

Upon achieving dynamic equilibrium, \(\dot{q} = 0\), the rate equation in Eq. \ref{EqS2} can thereupon be used to calculate the expected photon number. Defining the mode's equivalent temperature  as
\(T_{\text{mode}} = \frac{q}{\epsilon} = \frac{q h f_{\text{mode}}}{k_\text{B}}\), we obtain, for a single coupling port:
\begin{equation} \tag{S3} \label{EqS4}
    T_{\text{mode}} = \frac{T_0 + \kappa T_{\text{cold}}}{1 + \kappa}.
\end{equation}
Note that this steady-state solution of the photon-number dynamics is consistent with the application of Siegman's ``useful noise theorem'' (see chapter 8 of Ref.~\cite{siegman1964microwave}) as presented in the main text.   
\textit{In words}: the mode's expected temperature is the \textbf{weighted average} temperature of the two baths with which the mode interacts, the weighting of each bath being proportional to its interaction strength with the mode, \textit{i.e.}~its ``coupling''. Here, the coupling of the mode to the cavity's own intrinsic heat bath is (by definition/convention) unity.
Fig.~\ref{FS4} displays the mode temperature predicted by Eq. \ref{EqS4}. When the coupling factor of the over-coupling loop \(\kappa\) is large, the mode temperature \(T_{\text{mode}}\) will approach that of the cold load \(T_{\text{cold}}\).

\clearpage

\begin{figure}[!b]
\centering
%\figuretag{(S5)}
\includegraphics[width=0.65\columnwidth]{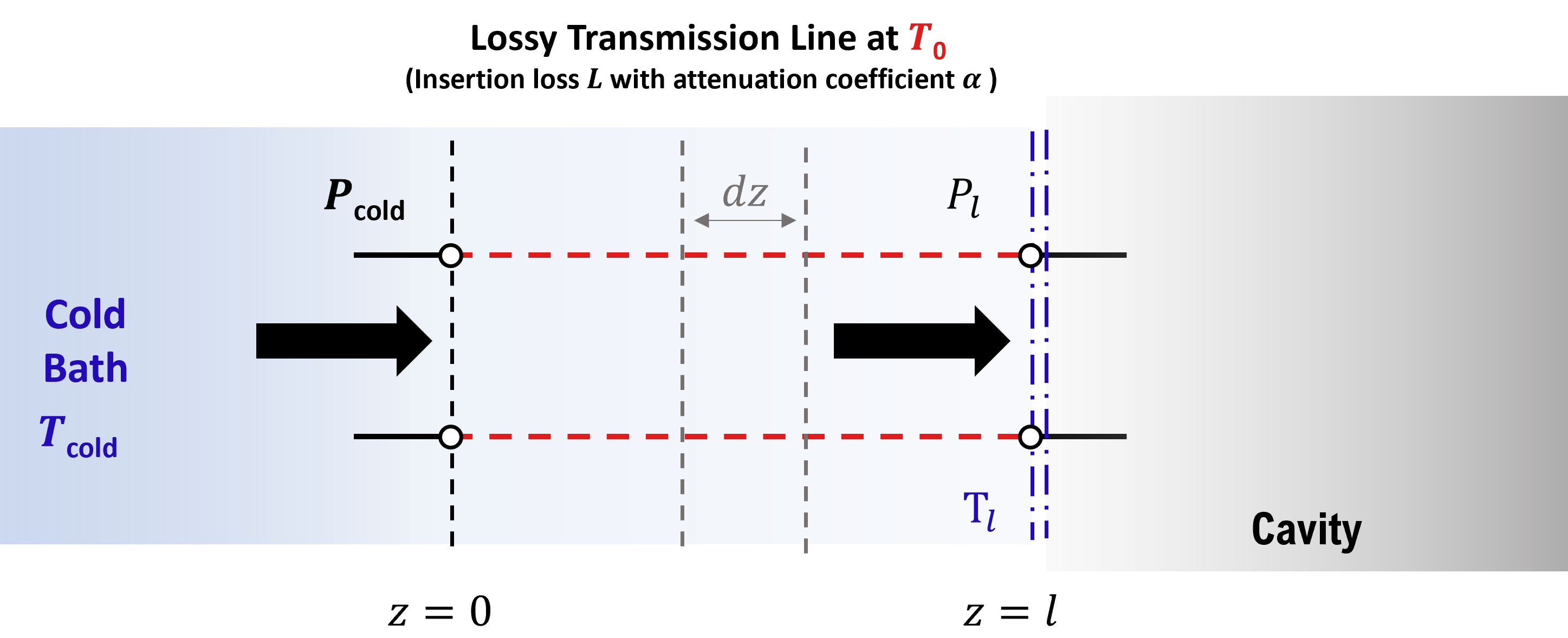}% Here is how to import EPS art
\caption{\label{FS5}
Calculating how the cooling LNA's noise temperature is degraded by transport along a length of (warm) lossy transmission line.}
\end{figure}

\textbf{\subsection{ Cavity cooling via a lossy transmission line}}
The input of a low-noise amplifier (LNA) can serve as an active cold load for reducing the thermal noise inside a cavity that is coupled to it. Here we analyse the additional thermal noise generated by a lossy transmission line, such as a length of coaxial cable, or any other lossy component inserted between the 
cold load and the cavity, as shown generically in Fig.~\ref{FS5}.
Again, see chapter 8 of Ref.~\cite{siegman1964microwave} for the conceptual
foundations and mathematics of the ``wave approach''\cite{djordjevic2017wave} used here. 
For initial simplicity, let's assume that the transmission line is uniform and held at a constant physical temperature \(T_0\) along its length.
The power \(P_{\text{sig}}\) of a signal flowing down such a line will decrease exponentially with distance obeying
\(dP_{\text{sig}}/dz = - 2 \alpha P_{\text{sig}}\), where \(\alpha\) is the line's voltage attenuation coefficient. Not only is the signal attenuated, but thermal noise is generated along the lossy line's length. At a distance \(z = l\) from the cold load, the forward-traveling noise power \(P_l\) will equal 
\begin{equation} \tag{S4} \label{EqS7}
    P_l = L P_{\text{cold}} + (1-L) P_{\text{line}},
\end{equation}
where \(P_{\text{cold}}\) and \(P_{\text{line}}\) are the available noise powers from a cold load at temperature \(T_{\text{cold}}\) and the end of an infinite length of the lossy transmission line at temperature \(T_0\), respectively; \(L\) is the insertion loss of the line between the cold load and the observation point at \(z = l\), where \(L = e^{-2 \alpha l}\) and \(L \leq 1\). Across a given frequency band of width \(B\), \( P_{\text{line}} = k_\text{B} T_{\text{line}} B\) (for \(hf/kT \ll 1\)).

In the case of a short transmission line (or any other 2-port component) with minimal insertion loss, such that
\( L \approx (1 -2 \alpha l) \) to good approximation, the equivalent noise power \(P\) at the end of the transmission line can be approximated (expressing the line's attenuation in decibels) as:

\begin{equation} \tag{S5} \label{EqS8}
    P_l = (1 - \frac{L_{\text{dB}}}{4.34}) P_{\text{cold}} + \frac{L_{\text{dB}}}{4.34} P_{\text{line}}.
\end{equation}
Note that this (slightly more complicated) approximation is more accurate than
equation (8-2-5) in Siegman's book \cite{siegman1964microwave}. 
Based on the relation between noise power and noise temperature, namely \( P = k_\text{B} T B \), the effective noise temperature of the cold load, upon including the additional noise from a lossy transmission line connected to it, can thereupon be approximated as:
\begin{equation} \tag{S6} \label{EqS9}
    T_l = (1 - \frac{L_{\text{dB}}}{4.34}) T_{\text{cold}} + \frac{L_{\text{dB}}}{4.34} T_0,
\end{equation}
where \(T_{\text{cold}}\) the temperature of the cold load itself
and \(T_0 = T_{\text{line}} \) is the (here assumed to be room) temperature of the transmission line.  
Combining Eq. \ref{EqS4} and Eq. \ref{EqS9}, the mode's noise temperature inside the microwave cavity is now given by:
\begin{equation} \tag{S7} \label{EqS10}
    T'_{\text{mode}} = \frac{T_0 + \kappa [(1 - \frac{L_{\text{dB}}}{4.34}) T_{\text{cold}} + \frac{L_{\text{dB}}}{4.34} T_{0}]}{1 + \kappa} = T_{\text{mode}}^{\text{orig.}}  + \Delta T_{\text{link}},
\end{equation}
where \( T_{\text{mode}}^{\text{orig.}} \) is the original mode temperature as given by Eq.~\ref{EqS4} above
and
\(\Delta T_{\text{link}} =  \frac{L_{\text{dB}}}{4.34} \kappa
(T_{0} - T_{\text{cold}})/(1 + \kappa)\) is the added noise temperature caused by the lossy transmission line.

\clearpage

\begin{figure}[htbp]
\centering
%\figuretag{(S6)}
\includegraphics[width=0.9\columnwidth]{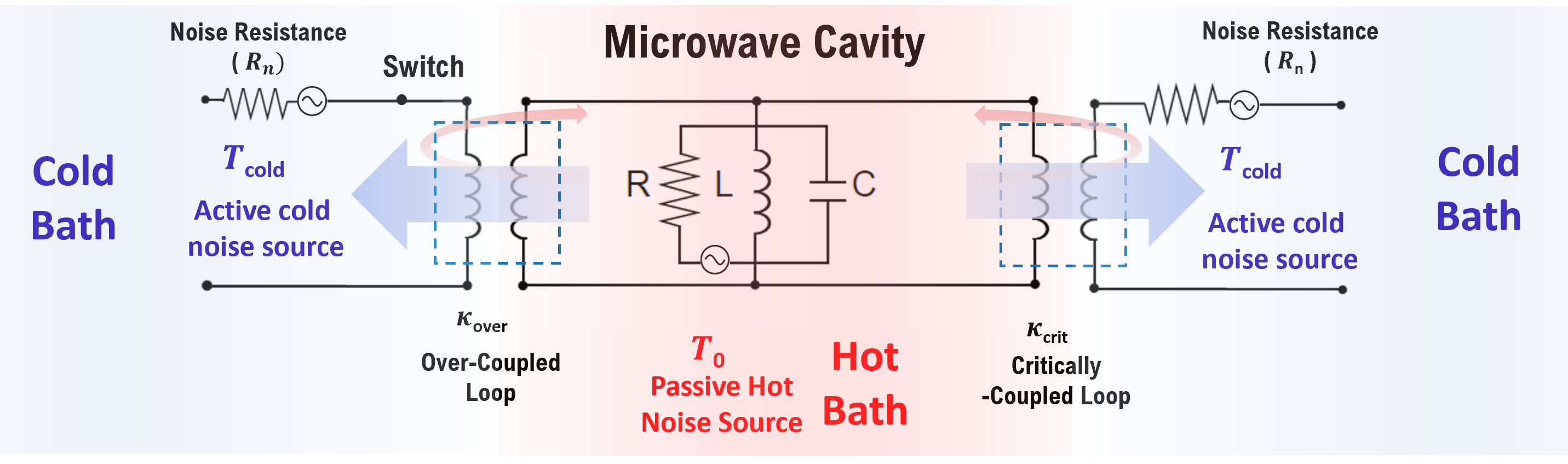}% Here is how to import EPS art
\caption{\label{FS6} 
Cavity cooling with two separate cold baths.
}
\end{figure}

\textbf{\subsection{Multiple cooling ports to a microwave cavity}}
We now consider a microwave cavity with two ports, one for cooling, the other for read-out, connected to separate LNAs functioning as cold loads.
With the setup depicted in Fig.~\ref{FS6} and applying Eq.~\ref{EqS2mo},
the rate equation governing the microwave cavity's photon number is:
\begin{equation} \tag{S8} \label{EqS11}
   \dot{q} = -  \underbrace{\omega_{\text{mode}} Q_0^{-1} (q - \epsilon T_0)}_{\colorbox{pink}{Hot bath}} -  \underbrace{\omega_{\text{mode}} (Q_0/\kappa_{\text{over}})^{-1} (q - \epsilon T_{\text{cold}}) }_{\colorbox{cyan}{Cold bath of cooling port}} -  \underbrace{\omega_{\text{mode}} (Q_0/\kappa_{\text{crit.}})^{-1} (q - \epsilon T_{\text{cold}}) }_{\colorbox{cyan}{Cold bath of monitoring port}},
\end{equation}
where \(\kappa_{over}\) and \(\kappa_{crit.}\) (\(\kappa_{crit.}  \equiv  1) \) are the coupling factors of the over-coupled and critically-coupled loops, respectively.
As in the previous analysis for a single port, the depth of cavity cooling upon reaching
dynamic equilibrium (\textit{i.e.}~``steady state'' conditions) is deduced by solving \(\dot{q} = 0\).
The mode temperature \(T_{\text{mode}}\) achieved, as appears in Fig.~\ref{FS6}, is given by:
\begin{equation} \tag{S9} \label{EqS12}
    T_{\text{mode}} = \frac{T_0 + (\kappa_{\text{over}} + \kappa_{\text{crit.}}) T_{\text{cold}}}{1 + \kappa_{\text{over}} + \kappa_{\text{crit.}}} = \frac{T_0 + (1 + \kappa_{\text{over}}) T_{\text{cold}}}{2 + \kappa_{\text{over}}}.
\end{equation}
In general, the temperature of a mode of a cavity with multiple cooling ports can be expressed as a weighted average:
\begin{equation} \tag{S10} \label{EqS13}
    T_{\text{mode}} = \frac{T_0 +  \sum_{i = 1}^n \kappa_i T_i }{1 + \sum_{i = 1}^n \kappa_i},
\end{equation}
where \(\kappa_i\) and \(T_i\) represent the coupling factor and noise temperature of the \(i\)-th cooling (or warming) element coupled to the mode,
respectively.

\clearpage
 \textbf{\subsection{Multiple cooling ports to a microwave cavity with lossy transmission lines}}
In this section, we consider a microwave cavity with multiple cooling ports and lossy transmission lines, integrating concepts from sections 2.3 and 2.4. Combining equations Eq.~\ref{EqS9} and Eq.~\ref{EqS13}, yields
\begin{equation} \tag{S11} \label{EqS14}
    T_{\text{mode}} = \frac{T_0 + \kappa_1[(1 - \frac{L_{\text{dB,1}}}{4.34}) T_{\text{cold}} + \frac{L_{\text{dB,1}}}{4.34}T_{0}] + \kappa_2 [(1 - \frac{L_{\text{dB,2}}}{4.34}) T_{\text{cold}} + \frac{L_{\text{dB,2}}}{4.34}T_{0}]}{1 + \kappa_1 + \kappa_2},
\end{equation}
where \(\kappa_1\) and \(\kappa_2\) represent the coupling coefficients of the first and second cold baths, respectively. We here assume the insertion losses of the transmission lines associated with the cold baths (\(L_{\text{dB,1}}\), \(L_{\text{dB,2}}\)) are small. If however the insertion loss of a connecting cable/device is not small, one needs to revert back to  
 Eq. \ref{EqS7}, equivalent to equation in (8-2-4) in Siegman's book \cite{siegman1964microwave}.
 Based on the relation between noise power and noise temperature, \(P = k_\text{B} T B\), the equivalent noise temperature of a cold load at temperature \(T_\text{cold}\), observed through a length of transmission line at temperature \(T_0\) and of significant linear insertion loss \( L = 10^{-L_{\text{dB}/10}}\), can be quantified as:
\begin{equation} \tag{S12} \label{EqS15}
\begin{split}
    T_l &= L T_{\text{cold}} + (1-L) T_{0} \\
    &= 10^{-\frac{L_{\text{dB}}}{10}} T_{\text{cold}} + \left(1 - 10^{-\frac{L_{\text{dB}}}{10}}\right) T_{0}.
\end{split}
\end{equation}
 Our experimental set-up includes two cold baths: one of them, namely the input of the cooling LNA, is connected to the cavity (when selected by the SPDT switch) via a small insertion loss, whereas the other heat bath, namely the monitoring LNA's input, is connected to the cavity via a large insertion loss. The most appropriate formula for calculating the temperature of the cavity's operational mode is thus:
\begin{equation} \tag{S13} \label{EqS16}
    T_\text{mode} = \frac{T_{0}+\kappa_1 [(1 - \frac{L_{\text{dB,1}}}{4.34}) T_{\text{cold}}+\frac{L_{\text{dB,1}}}{4.34}T_{0}]+\kappa_2 [10^{-\frac{L_{\text{dB,2}}}{10}} T_{\text{cold}} + (1 - 10^{-\frac{L_{\text{dB,2}}}{10}})  T_{0}]}{1+\kappa_1+\kappa_2}.
\end{equation}
These approximations are analyzed in Fig.~\ref{FS7} below. 

\begin{figure}[b]
\centering
%\figuretag{(S7)}
\includegraphics[width=0.75\columnwidth]{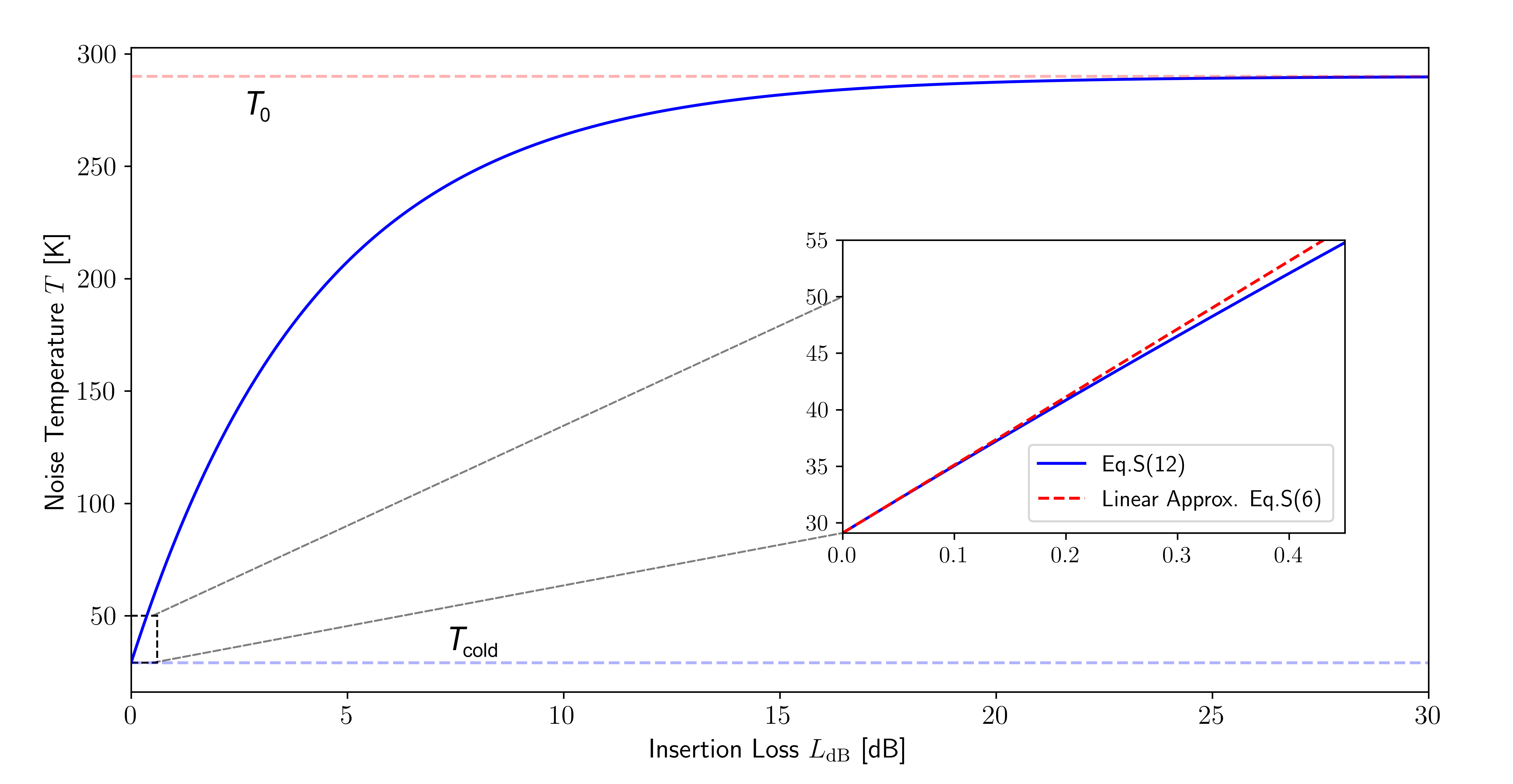}
\caption{\label{FS7}
How, according to Eq.~\ref{EqS15}, the forward noise power (in units of absolute temperature \(T\)) at the far end of a lossy cable depends on the cable's insertion loss \(L_{\text{dB}}\). Here
the cable is uniformly maintained at a temperature of \(T_{\text{0}}=290\)~K throughout its length and
the near end of the cable is terminated and cooled by the ACL whose noise temperature is \(T_{\text{cold}}=29.1\)~K (dashed cornflower-blue baseline)--corresponding to the input noise temperature of
the Qorvo QPL9547EVB-01 LNA used experimentally.
The zoomed-in area details how, when the insertion loss is less than 0.5~dB,
equation Eq.~\ref{EqS9} (red dashed line -``Linear Approx.''), \textit{i.e.}~the linear approximation  provided by equation (8-2-5) in Siegman's book \cite{siegman1964microwave}, compares against the exact formula Eq.~\ref{EqS15} (blue solid line).
}
\end{figure}

%\clearpage

\textbf{\section{Noise Analysis}} \label{sec:NA}

Similar to the previous work by Wu et al. \citep{wu2021bench}, we also adopt the ``wave approach'' (see Chapter 8 of Siegman \citep{siegman1964microwave}) to analyze the flow of noise power from the cavity into our homodyne receiver, as shown in Fig.~\ref{FS8}. Equation S8 in the supplemental material of Wu et al.'s paper provides a formula for the factor,
\(\Delta P\) (in units of dB), by which the measured noise power is reduced by mode cooling:
\begin{equation} \tag{S14} \label{EqS17o}
\Delta P'  = 10 \log_{10} \frac{  {G_\text{LNA}[(T_\text{min} + T_\text{mode}) (1-|\Gamma_c|^2) + 4 T_{0} \frac{R_\text{n}}{Z_{0}} \frac{\lvert \Gamma_\text{c} -\Gamma_\text{opt} \rvert^2}{\lvert 1+\Gamma_\text{opt} \rvert^2} +T_\text{image} ]+T_\text{REC}}  }{  {G_\text{LNA}[(T_\text{min} + T_\text{mode}^0) (1-|\Gamma_c^0|^2) + 4 T_{0} \frac{R_\text{n}}{Z_\text{0}} \frac{\lvert \Gamma_c^0 -\Gamma_\text{opt} \rvert^2}{\lvert 1+\Gamma_\text{opt} \rvert^2} +T_\text{image} ]+T_\text{REC}}  },
\end{equation}
However, this formula needs to be modified to suit our experimental set-up, and for the following reasons:
(i) Our current receiver is a homodyne, not a single-conversion superhet subject to additional image noise;
one can thus set \(T_\text{image} = 0\). (ii) Even when monitoring (with the cooling LNA disconnected), the cavity mode is continuously cooled (albeit only slightly) by the receiver's front-end LNA, connected to the cavity's monitoring port via a lossy stub tuner and cable (substantially reducing the size of the the LNA's cooling effect) --see main text.
The ambient/reference mode temperature in the above formula, namely \(T_\text{mode}^0\), should thus be set equal (again, see main text) to \(T_\text{mode}^\text{ambient}\approx\)~257.6~K, somewhat less than~290~K.
Finally, (iii) the ambient (reference) and post-cooled measurements are performed with the cavity under the same coupling conditions, namely with a single, critically-coupled monitoring port; the reflection amplitude off
this port is always zero (during measurement of the noise power); thus, one can set \(\Gamma_c = \Gamma_c^0 = 0\).
These largely simplifying modifications lead to:
\begin{equation} \tag{S15} \label{EqS17}
\Delta P  = 10 \log_{10} \frac{  {G_\text{LNA}[(T_\text{min} + T_\text{mode}) + 4 T_{0} \frac{R_\text{n}}{Z_{0}} \frac{\lvert \Gamma_\text{opt} \rvert^2}{\lvert 1+\Gamma_\text{opt} \rvert^2} ]+T_\text{REC}}  }
{  {G_\text{LNA}[(T_\text{min} + T_\text{mode}^\text{ambient}) + 4 T_{0} \frac{R_\text{n}}{Z_\text{0}} \frac{\lvert \Gamma_\text{opt} \rvert^2}{\lvert 1+\Gamma_\text{opt} \rvert^2} ]+T_\text{REC}}  },
\end{equation}
where
\(T_\text{mode}\) is the pre-cooled mode temperature; \(T_\text{min}\), \(R_\text{n}\), \(\Gamma_\text{opt}\)
are the four noise parameters (being a complex number, \(\Gamma_\text{opt}\) counts double) of the monitoring LNA,
namely a Qorvo QPL9547EVB-01; the reference impedance \(Z_0 = 50\)~\( \Omega\) throughout.
The values assigned to these parameters, as shown in Table~\ref{T1},
have been interpolated to 1.45~GHz from measurements at straddling spot
frequencies found in the manufacturer's data sheet \cite{QPL9547}.
The noise temperature of the rest of the receiver, treated as a cascade of amplifiers, was
calculated using Friis' formula and
amplifiers' specified noise temperatures and/or equivalent noise figures;
see Table~\ref{T1}. The monitoring LNA's gain, \(G_\text{LNA}\),
was measured using a VNA and found to be consistent with the QPL9547's datasheet.
The amplitude of the applied interrogating microwave tone was chosen to be sufficiently low that stimulated emission/absorption is very slow compared to the intrinsic spin dynamics of pentacene's optically excited triplet state. 
\begin{figure}[!b]
 \centering
% \figuretag{(S8)}
\includegraphics[width=0.9\columnwidth]{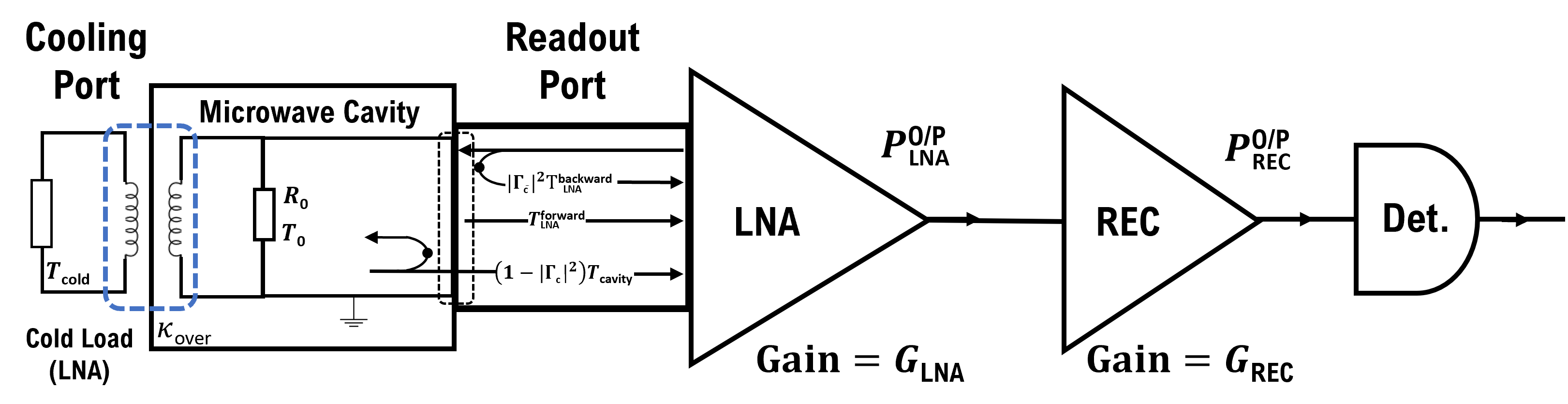}% Here is how to import EPS art
\caption{ \label{FS8}
Anatomy of the signal path at the front end of the homodyne receiver, displaying the forward
and backward noise powers between the LNA and microwave cavity.}
\end{figure}
%\clearpage
\begin{table}[htpb]
\centering
\caption{Values of parameters -entering the noise-reduction equation \ref{EqS17}}
\label{T1}
\begin{tabular}{lcc}
\hline \hline
Description of the Receiver                   & Symbol                  & Value (@ 1.4495 GHz)        \\ \hline
Linear power gain of first LNA                & \(G_{\text{LNA}}\)      & 166.0 $\pm 3.3$         \\
Minimum noise figure of first LNA             & \(NF_{\text{min}}\)     & 0.17*                  \\
Minimum noise temperature of first LNA        & \(T_{\text{min}}\)      & 11.6 K*                 \\
Optimum source reflection coefficient         & \(\Gamma_{\text{opt}}\) & (0.073+0.125i)*        \\
Noise resistance of first LNA                 & \(R_\text{n}\)          & 2.00 $\Omega$*          \\
Reflection looking into input of first LNA    & \(S_{11}\)              & -(0.107+0.234i)*       \\
Noise temperature of the rest of the receiver & \(T_{\text{REC}}\)      & 36.1 K$^\dagger$        \\ \hline \hline
* interpolated from published datasheet \cite{QPL9547}\\
$\dagger$ calculated using Friis' formula \cite{friis44}
\end{tabular}
\end{table}

Given the parameters appearing in Table~\ref{T1}, whose values
compose the noise calibration of our tr-EPR spectrometer,
the observed reduction in noise power provided by precooling the cavity, as predicted by Eq.~\ref{EqS17},
depends on the value \(T_{\text{mode}}\) in the numerator versus that of
\(T_\text{mode}^\text{ambient}\) in the denominator. 
The latter, corresponding to when the cavity's \( \textrm{TE}_{01 \delta} \) mode
is critically coupled to its monitoring LNA via a lossy stub tuner, should equal
\((T_{0} + T_{\text{mon.}})/2 = 257.6\)~K, where \(T_{\text{mon.}} = 10^{-\frac{L_{\text{dB,2}}}{10}} T_{\text{cold}}+(1 - 10^{-\frac{L_{\text{dB,2}}}{10}}) T_{0} \approx 225.2\) K according to Eq. \ref{EqS15}.
Upon fixing the value of \(T_\text{mode}^\text{ambient}\), Eq.~\ref{EqS17}
can be used to infer \(T_{\text{mode}}\) from the experimentally observed 
reduction in noise power \(\Delta P  \). 
Using the data shown in Fig.~8(b), a rough estimate of \(\Delta P \) can be arrived
at from the difference in the spectral noise power, averaged from 5 to 10 MHz (so avoiding the 1/f noise prevalent below 1 MHz), between ``pre-cooled'' and ``ambient'' conditions;
one obtains: \(\Delta P_\text{exp-EPR} = -3.7 \pm 1\)~dB. Since this estimate of \(\Delta P\) has greater uncertainty than that derived directly from monitoring the pre-cooling in the time domain, namely  \(\Delta P_\text{exp-Cool} \), it is not further used.

The mode temperature \(T_{\text{mode}}\) can be separately predicted using our model, namely Eq.~\ref{EqS16} above; it depends on the (presumed identical) noise temperature of the two LNAs, \(T_{\text{cold}}\), the coupling factor of the over-coupling loop \(\kappa_{\text{over}}\), and the insertion losses of the RF switch \(L_{\text{dB, 1}}\) and the monitoring LNA's connection to the cavity line \(L_{\text{dB,2}}\).
One arrives at
\begin{equation} \tag{S16} \label{EqS18}
T_{\text{mode}} = \frac{T_{\text{0}}+\kappa'_{\text{over}}[(1- \frac{L_{\text{dB,1}}}{4.34})T_{\text{cold}}+\frac{L_{\text{dB,1}}}{4.34}T_{0}]+[10^{-\frac{L_{\text{dB,2}}}{10}} T_{\text{cold}} + (1 - 10^{-\frac{L_{\text{dB,2}}}{10}})  T_0]}{2+\kappa'_{\text{over}}} \approx 131.5  \pm 5~\text{K}.
\end{equation}
The values of \(L_{\text{dB,1}}\) and \(L_{\text{dB,2}}\) at 1.4495 GHz were measured using
our VNA (HP-8753C) to be \(0.19 \pm 0.01 \)~dB and \(6.05 \pm 0.01 \)~dB, respectively. The coupling factor between the cavity and the cold load was characterized by the method(s) of Kajfez and Hwan \cite{kajfez1984q} using the same VNA; we obtained \(\kappa_{\text{over}} =3.8 \pm 0.3\). The value of \(T_{\text{cold}}\) is calculated to be \(29.1\pm 0.5\)~K based on the definition of noise parameters \cite{escotte1993evaluation,escotte2012}
utilizing the parameters listed in Table~\ref{T1}. \(T_0\) represents room temperature, set at \(290\)~K. The uncertainty of \(T_{\text{mode}}\) shown in Eq.~\ref{EqS18} is calculated by applying Pythagoras' theorem to error propagation.  Based on this result, we can use Eq.~\ref{EqS17} to obtain \(\Delta P_{\text{model}} = -3.2 \pm 0.2\)~dB.

\begin{figure}[htpb]
\centering
%\figuretag{(S9)}
\includegraphics[width=0.85\columnwidth]{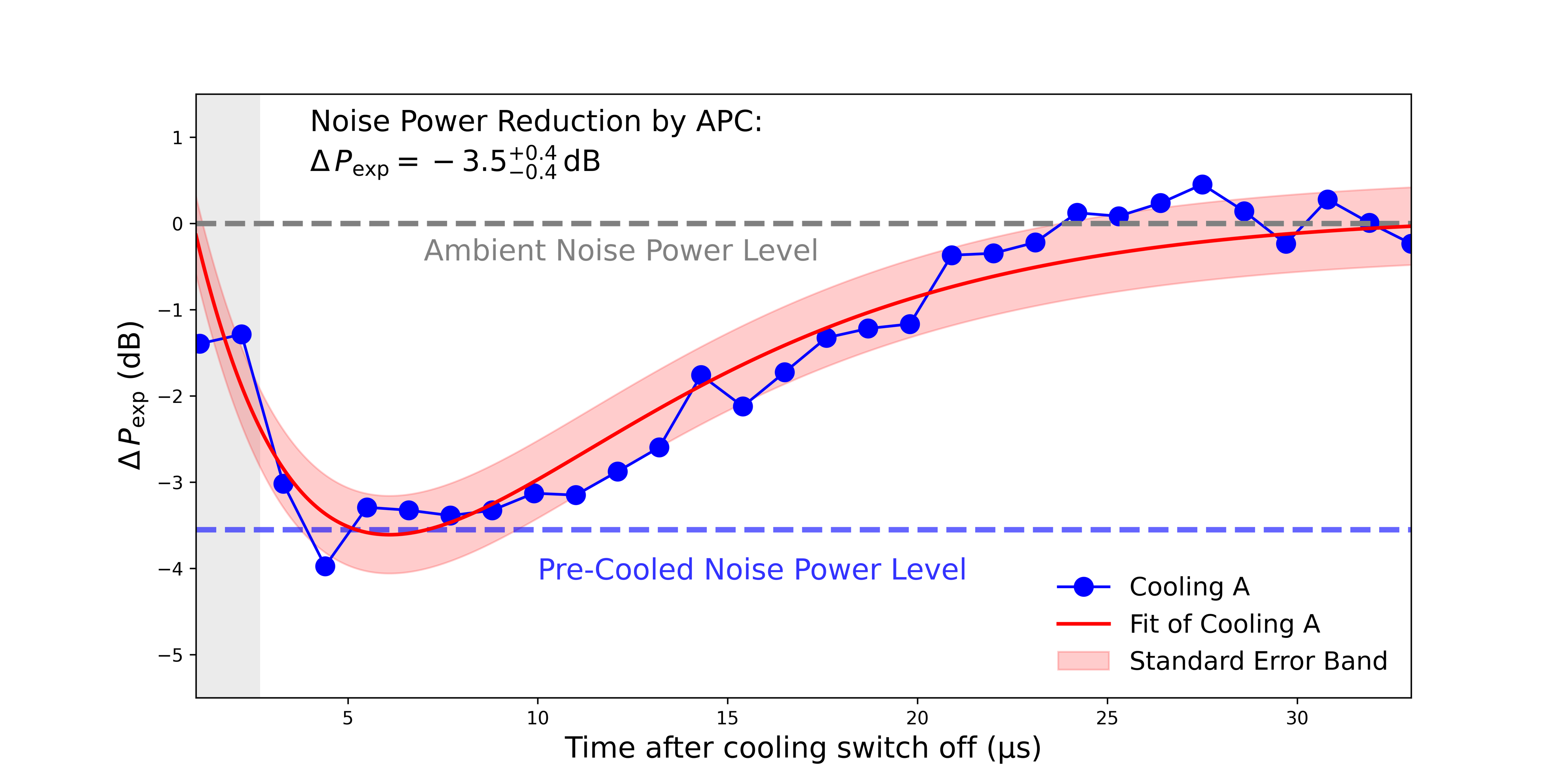}% Here is how to import EPS art
\caption{ \label{FS9}
Error analysis and fitting of the experimental warming-up signal as presented in Fig.~6 of the main paper. The grey area at shorts times (after switching off the cooling LNA) on the far left represents the period when the double-throw switch is still transitioning to connect the over-coupled port to an ``open'' load, disconnecting the ACL from the resonator; this takes around $t_{\text{switch}} = 2$~$\mu$s.  Each blue dot plots \(\Delta P_{\text{exp}} = 10 \log_{10}
(< V^2_\text{cold} > /
< V^2_\text{ambient} >) \) in
units of dB\cite{sung2003transient}, where 
\( < V^2_\text{cold} >\) is the mean-squared received noise voltage averaged over ten consecutive data points, spaced 100 ns apart, spanning one microsecond; \( < V^2_\text{ambient} >\) is the equivalent mean-squared noise voltage under ambient conditions (based on an adjoining sample spanning several hundred microseconds).
The solid red line, along with its surrounding salmon pink standard error band, was obtained by least-squares fitting a bi-exponential to the evolution of \(\Delta P_{\text{exp}} \) excluding the initial switching period.
The longer of the two decay constants was determined to be \(\tau_{\text{warm up}}^{\text{exp.}} \approx 9 \pm 1 \mu\)s.
}
\end{figure}

The depth of mode cooling was also estimated by fitting a bi-exponential function
to the increase in measured noise power upon switching out the cooling LNA, as shown in Figs.~4(c) and 7  in the main paper. The best-fit curve, with its associated standard error band, is depicted in Fig.~\ref{FS9}; this determines \(\Delta P_{\text{exp-Cool}}= -3.5 \pm 0.4\)~dB, agreeing (within error bars) of both our model's prediction \(\Delta P_{\text{model}}\) and our analysis using the noise spectrum from tr-EPR data, namely \(\Delta P_{\text{exp-EPR}}\) as shown in Fig.~8(b) of the main text. Our estimates are summarized in Table~\ref{table:comparison}.

\begin{table}[ht]
\centering
\caption{Modeled and experimentally-inferred depth of cooling}
\begin{tabular}{rccc}
\toprule
\toprule
 method:& \multicolumn{1}{c}{Model} & \multicolumn{1}{c}{Direct monitoring} & \multicolumn{1}{c}{EPR data analysis} \\
 & \multicolumn{1}{c}{} & \multicolumn{1}{c}{(time-domain)} & \multicolumn{1}{c}{(frequency-domain)} \\
 label:
 & \multicolumn{1}{c}{model} & \multicolumn{1}{c}{exp-Cool} & \multicolumn{1}{c}{exp-EPR} \\
\midrule
Noise-power reduction (\(\Delta P\) ) in dB: & 3.2 & 3.5 & 3.7 \\
Standard deviation of \(\Delta P \)  in dB: & $\pm$ 0.2 & $\pm$ 0.4 & $\pm$ 1.0 \\
Inferred mode temperature ($T$\textsubscript{mode}) in K & 131.5 & 122.6 & - \\
Error in $T$\textsubscript{mode} in K & \{+7,-6\} & \{+13, -12\} & - \\
\bottomrule
\bottomrule
\end{tabular}
\label{table:comparison}
\end{table}

\clearpage

\nocite{*}

%\bibliography{sample631}% Produces the bibliography via BibTeX.

%\bibliographystyle{ieeetr}



\end{document}